\documentclass[letterpaper,twocolumn,10pt]{article}
\usepackage{usenix,epsfig,endnotes,hyphenat,subfig,inputenc,float,xcolor}
\usepackage[titletoc,title]{appendix}
\usepackage{listings}
\usepackage{textcomp}
\usepackage{float}
\usepackage{color}
\usepackage[hyphens]{url}
\usepackage{graphicx}
\usepackage{multirow}
\usepackage{array}
\usepackage[font=small,format=plain,labelfont=bf,textfont=bf]{caption}
\usepackage{mdwlist}
\usepackage{booktabs}
\usepackage{listings}
\usepackage{enumitem}
\usepackage{xspace}
\usepackage{hyperref}
\usepackage{multicol}
\usepackage{stfloats}
\usepackage{tikz}
\usepackage[small,compact]{titlesec}
\newenvironment{Figure}
  {\par\medskip\noindent\minipage{\linewidth}}
  {\endminipage\par\medskip}
\definecolor{customBlue}{rgb}{0.27,0.45,0.77}
\definecolor{heraldBlue}{rgb}{0.0,0.0,0.8}
\definecolor{heraldRed}{rgb}{0.8,0.0,0.0}
\definecolor{heraldGray}{rgb}{0.2,0.2,0.5}
\definecolor{heraldGreen}{rgb}{0.0,0.4,0.0}
\definecolor{heraldPink}{rgb}{0.8,0.1,0.6}
\definecolor{darkgray}{rgb}{0.3, 0.3, 0.3}

\newcommand{\ie}{i.e.,\xspace}
\newcommand{\eg}{e.g.,\xspace}
\newcommand{\etc}{etc.}

\newcommand{\smartparagraph}[1]{\noindent {\bf #1}}

\newcommand{\sys}[0]{RedN\xspace}
\newcommand{\Sys}[0]{\sys}

\newcommand{\sysseq}[0]{\sys-Seq\xspace}
\newcommand{\syspar}[0]{\sys-Parallel\xspace}

\newcommand{\rnoop}{\textsc{Noop}\xspace}
\newcommand{\rwrite}{\textsc{Write}\xspace}
\newcommand{\rread}{\textsc{Read}\xspace}
\newcommand{\rwriteimm}{\textsc{WriteImm}\xspace}

\newcommand{\rsend}{\textsc{Send}\xspace}
\newcommand{\rrecv}{\textsc{Recv}\xspace}
\newcommand{\rwait}{\textsc{Wait}\xspace}
\newcommand{\renable}{\textsc{Enable}\xspace}
\newcommand{\rcas}{\textsc{CAS}\xspace}
\newcommand{\radd}{\textsc{ADD}\xspace}

\newcommand{\rmin}{\textsc{Min}\xspace}
\newcommand{\rmax}{\textsc{Max}\xspace}

\newcommand{\us}{{\textmu}s\xspace}

\DeclareRobustCommand{\circledblack}[1]{\tikz[baseline=(char.base)]{
            \node[shape=circle,fill=black,inner sep=0pt, minimum size=12pt,scale=0.9] (char) {\textcolor{white}{#1}};}}

\DeclareRobustCommand{\circledblue}[1]{\tikz[baseline=(char.base)]{
            \node[shape=circle,fill=customBlue,inner sep=0pt, minimum size=12pt,scale=0.9] (char) {\textcolor{white}{#1}};}}

\DeclareRobustCommand{\circledgray}[1]{\tikz[baseline=(char.base)]{
            \node[shape=circle,fill=darkgray,inner sep=0pt, minimum size=12pt,scale=0.9] (char) {\textcolor{white}{#1}};}}

\usepackage[page]{totalcount}
\begin{document}

\date{}

\title{\Large \bf RDMA is Turing complete, we just did not know it yet!}

\author{
{\rm Waleed Reda}\\
Universit\'{e} catholique de Louvain \\ KTH Royal Institute of Technology
\and
{\rm Marco Canini}\\
KAUST
\and
{\rm Dejan Kosti\'{c}}\\
KTH Royal Institute of Technology
\and
{\rm Simon Peter}\\
UT Austin
}

\maketitle

\begin{abstract}
\addcontentsline{toc}{section}{Abstract}

It is becoming increasingly popular for distributed systems to exploit
offload to reduce load on the CPU. Remote Direct Memory
Access (RDMA) offload, in particular, has become popular. However, RDMA still requires CPU intervention for complex offloads that go beyond simple remote memory
access. As such, the offload potential is limited and
RDMA-based systems usually have to work around such limitations.

We present \sys, a principled, practical approach to implementing
complex RDMA offloads, without requiring any hardware
modifications. Using \emph{self-modifying} RDMA chains, we lift the
existing RDMA verbs interface to a Turing complete set of programming
abstractions. We explore what is possible in terms of offload
complexity and performance with a commodity RDMA NIC. We show how to
integrate these RDMA chains into applications, such as the Memcached
key-value store, allowing us to offload complex tasks such as key
lookups. \sys can reduce the latency of key-value get operations by up
to 2.6$\times$ compared to state-of-the-art KV designs that use
one-sided RDMA primitives (\eg FaRM-KV), as well as traditional
RPC-over-RDMA approaches. Moreover, compared to these baselines, \sys
provides performance isolation and, in the presence of contention, can
reduce latency by up to 35$\times$ while providing applications with
failure resiliency to OS and process crashes.

\end{abstract}

\section{Introduction}
\label{sec:intro}

As server CPU cycles become an increasingly scarce resource, offload is
gaining in popularity
\cite{nica,hyperloop,kvdirect,ipipe,e3,floem}. System operators wish
to reserve CPU cycles for application execution, while common,
oft-repeated operations may be offloaded. NIC offloads,
in particular, have the benefit that they reside in the network data
path and NICs can carry out operations on in-flight data with low
latency \cite{ipipe}.

For this reason, remote direct memory access (RDMA) \cite{rdma_spec}
has become ubiquitous~\cite{ibmarket}. Mellanox ConnectX NICs \cite{connectx} have
pioneered ubiquitous RDMA support and Intel has added RDMA support to
their 800 series of Ethernet network adapters \cite{intel800}. RDMA
focuses on the offload of simple message passing (via {\rsend}/{\rrecv} verbs)
and remote memory access (via {\rread}/{\rwrite} verbs)
\cite{rdma_spec}. Both primitives are widely used in networked
applications and their offload is extremely useful. However, RDMA is
not designed for more complex offloads that are also common in
networked applications. For example, remote data structure
traversal and hash table access are not normally deemed realizable
with RDMA~\cite{strom}. This led to many RDMA-based systems requiring multiple
network round-trips or to reintroduce involvement of the server's CPU
to execute such requests~\cite{aguilera2019impact,farm,herd,kazhamiaka2019sift,pilaf, poke2015dare, wang2017apus}.

To support complex offloads, the networking community has developed a
number of SmartNIC architectures
\cite{catapult,xpliant,bluefield,netfpga,stingray}. SmartNICs
incorporate more powerful compute capabilities via CPUs or FPGAs. They
can execute arbitrary programs on the NIC, including complex
offloads. However, these SmartNICs are not ubiquitous and their
smaller volume implies a higher cost. SmartNICs can cost up to
5.7$\times$ more than commodity RDMA NICs (RNICs) at the same link
speed (\S\ref{sec:smartnics}). Due to their custom architecture, they
are also a management burden to the system operator, who has to
support SmartNICs apart from the rest of the fleet.

We ask whether we can avoid this tradeoff and attempt to use the
ubiquitous RNICs to realize complex offloads. To do so, we have to
solve a number of challenges. First, we have to answer if and how we
can use the RNIC interface, which consists only of simple data
movement verbs ({\rread}, {\rwrite}, {\rsend}, {\rrecv}, etc.) and no
conditionals or loops, to realize complex offloads. Our solution has
to be general so that offload developers can use it to build complex
\emph{RDMA programs} that can
perform a wide range of functionality. Second, we have to ensure that
our solution is efficient and that we understand the performance and
performance variability properties of using RNICs for complex
offloads. Finally, we have to answer how complex RNIC offloads
integrate with existing applications.

In this paper, we show that RDMA is \emph{Turing complete}, making it possible to use RNICs to
implement complex offloads. To do so, we implement conditional
branching via \emph{self-modifying} RDMA verbs. Clever use of the
existing compare-and-swap ({\rcas}) verb enables us to dynamically
modify the RNIC execution path by editing subsequent verbs in an RDMA
program, using the {\rcas} operands as a predicate. Just like
self-modifying code executing on CPUs, self-modifying verbs require
careful control of the execution path to avoid consistency issues due
to RNIC verb prefetching. To do so, we rely on the {\rwait} and
{\renable} RDMA verbs~\cite{mellanox-dev-guide, hyperloop} that provide execution dependencies. {\rwait} allows us to halt
execution of new verbs until past verbs have completed, providing
strict ordering among RDMA verbs. By controlling verb prefetching,
{\renable} enforces consistency for verbs modified by preceding
verbs. {\renable} also allows us to create loops by re-triggering
earlier, already-executed verbs in an RDMA work queue---allowing
the NIC to operate autonomously without CPU intervention.

Based on these primitives, we present \sys, a principled, practical
approach to implementing complex RNIC offloads. Using self-modifying
RDMA programs, we develop a number of building blocks that lift the existing RDMA
verbs interface to a Turing complete set of programming abstractions. Using these
abstractions, we explore what is possible in terms of offload complexity and
performance with just a commodity RNIC. We show how to integrate complex RNIC
offloads, developed with \sys principles, into existing networked applications.
\sys affords offload developers a practical way to implement complex NIC offloads
on commodity RNICs, without the burden of acquiring and maintaining SmartNICs. Our
code is available at: \url{https://redn.io}. 

We make the following contributions:

\noindent $\bullet$ We present \sys, a principled, practical approach to offloading
  arbitrary computation to RDMA NICs. \sys leverages RDMA ordering and
  compare-and-swap primitives to build conditionals and loops. We show that these primitives are sufficient to make RDMA Turing complete.

\noindent $\bullet$ Using \sys, we present and evaluate the implementation of various
  offloads that are useful in common server computing scenarios. In
  particular, we implement hash table lookup with Hopscotch hashing
  and linked list traversal.%

  \noindent $\bullet$ We evaluate the complexity and performance of
  offload in a number of use cases with the Memcached key-value
  store. In particular, we evaluate offload of common key-value get
  operations, as well as performance isolation and failure resiliency
  benefits. We demonstrate that RNIC offload with \sys can realize all
  of these benefits. It can reduce average latency of get operations
  by up to 2.6$\times$ compared to state-of-the-art one-sided RDMA
  key-value stores (\eg FaRM-KV~\cite{farm}), as well as traditional
  two-sided RPC-over-RDMA implementations. Moreover, \sys provides
  superior performance isolation, improving latency by up to
  35$\times$ under contention, while also providing higher
  availability under host-side failures.

\section{Background}

RDMA was conceived for high-performance computing (HPC) clusters, but
it has grown out of this niche~\cite{ibmarket}. It is becoming
ever-more popular due to the growth in network bandwidth, with
stagnating growth in CPU performance, making CPU cycles an
increasingly scarce resource that is best reserved to running
application code.  With RNICs now considered commodity, it is
opportunistic to explore the use-cases where their hardware can yield
benefits. These efforts, however, have been limited by the RDMA API,
which constrains the expression of many complex
offloads. Consequently, the networking community has built SmartNICs
using FPGAs and CPUs to investigate new complex offloads.

\subsection{SmartNICs}
\label{sec:smartnics}
To enable complex network offloads, SmartNICs have been
developed~\cite{agiliocx,bluefield,liquidio,catapult}. SmartNICs
include dedicated computing units or FPGAs, memory, and several
dedicated accelerators, such as cryptography engines. For example,
Mellanox BlueField~\cite{bluefield} has 8$\times$ARMv8 cores with
16GB of memory and 2$\times$25GbE ports. %
These SmartNICs are capable of running full-fledged operating systems, but also ship
with lightweight runtime systems that can provide kernel-bypass access
to the NIC's IO engines.

\smartparagraph{Related work on SmartNIC offload.} SmartNICs have
been used to offload complex tasks from server CPUs. For example,
StRoM~\cite{strom} uses an FPGA NIC to implement RDMA verbs and
creates generic kernels (or building blocks) that perform various
functions, such as traversing linked lists. KV-Direct~\cite{kvdirect}
uses an FPGA NIC to accelerate key-value
accesses. iPipe~\cite{ipipe} and Floem~\cite{floem} are programming
frameworks that simplify complex offload development for primarily
CPU-based SmartNICs. E3~\cite{e3} transparently offloads microservices
to SmartNICs.

\smartparagraph{The cost of SmartNICs.} While SmartNICs provide the
capabilities for complex offloads, they come at a cost. For example, a
dual-port 25GbE BlueField SmartNIC at \$2,340 costs 5.7$\times$ more
than the same-speed ConnectX-5 RNIC at \$410
(cf.~\cite{mellanox_store}).
Another cost is the additional management required for
SmartNICs. SmartNICs are a special piece of complex equipment that
system administrators need to understand and maintain. SmartNIC
operating systems and runtimes can crash, have security flaws, and
need to be kept up-to-date with the latest vendor patches. This is an
additional maintenance burden on operators that is not incurred by
RNICs.

\subsection{RDMA NICs}

The processing power of RDMA NICs (RNICs) has doubled with each subsequent
generation. This allows RNICs to cope with higher packet rates and
more complex, hard-coded offloads (e.g., reduction operations,
encryption, erasure coding). 

We measure the verb processing bandwidth of several generations of Mellanox ConnectX NICs,
using the Mellanox \verb+ib_write_bw+ benchmark. This benchmark
performs 64B RDMA writes and, as such, it is not network bandwidth limited due to
the small RDMA write size. We find that the verb processing bandwidth
doubles with each generation, as we can see in
Table~\ref{table:cx_performance}. This is primarily due to a doubling
in processing units (PUs) in each generation.\footnote{Discussions with Mellanox affirmed our findings.}
As a result, ConnectX-6 NICs can execute up to 110 million RDMA verbs per
second using a single NIC port. This increased hardware performance
further motivates the need for exploiting the computational power of
these devices.

\begin{table}[b]
\centering
\resizebox{0.85\columnwidth}{!}{%
\tiny
\begin{tabular}{|c|c|c|}
\hline
\textbf{RNIC} & \textbf{PUs} & \textbf{Throughput} \\ \hline
ConnectX-3 (2014)              & 2                             & 15M verbs/s                          \\ \hline
ConnectX-5 (2016)              & 8                             & 63M verbs/s                          \\ \hline
ConnectX-6 (2017)              & 16                            & 112M verbs/s                         \\ \hline
\end{tabular}
}
\caption{Number of Processing Units (PUs) and performance of various ConnectX generations.}
\label{table:cx_performance}

\end{table}

\begin{figure*}[t]
    \centering
    \includegraphics[width=0.9\textwidth]{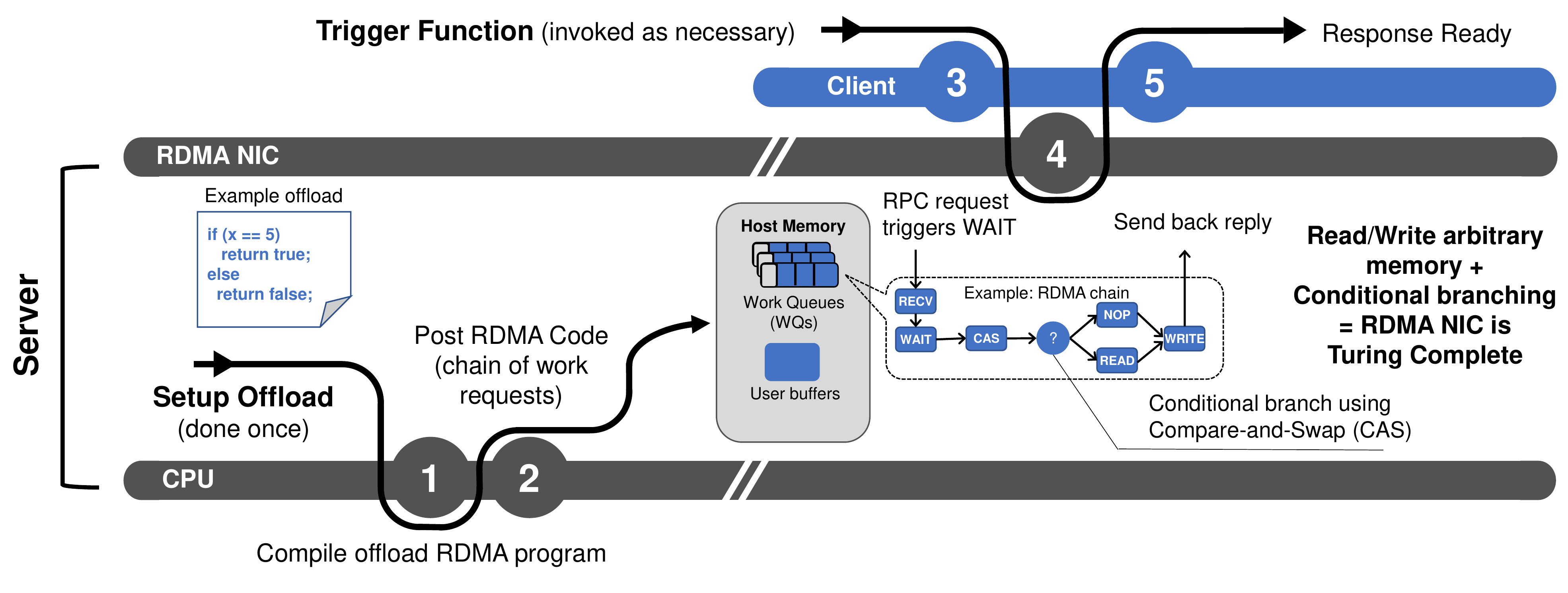}
    \caption{RDMA NICs can implement complex offloads if we allow conditional branches
      to be expressed. Conditional branching can be implemented by
      using CAS verbs to modify
      subsequent verbs in the chain, without any
      hardware modification.
        }
    \label{fig:turing_complete}
\end{figure*}

\smartparagraph{Related work on RDMA offload.} RDMA has been employed
in many different contexts, including accelerating key-value stores
and filesystems ~\cite{assise,farm,herd,pilaf,orion},
consensus~\cite{aguilera2019impact, kazhamiaka2019sift, wang2017apus,
  poke2015dare}, distributed locking~\cite{dslr}, and even nuanced
use-cases such as efficient access in distributed tree-based indexing
structures~\cite{ziegler2019designing}. These systems operate within
the confines of RDMA's intended use as a \emph{data movement} offload
(via remote memory access and message passing). When complex
functionality is required, these systems involve multiple RDMA
round-trips and/or rely on host CPUs to carry out the complex
operations.

Within the storage context, Hyperloop~\cite{hyperloop} demonstrated that pushing the RNIC offload capabilities is possible. Hyperloop combines RDMA verbs to
implement complex storage operations, such as chain
replication, without CPU involvement. However, it does not provide a blueprint for
offloading arbitrary processing and cannot offload functionality
that uses any type of conditional logic (e.g., walking a remote
data structure). Moreover, the Hyperloop protocol is likely incompatible
with next-generation RNICs, as its implementation relies
on changing work request ownership---a feature that is deprecated for
ConnectX-4 and newer cards.

Unlike this body of previous work, we aim to unlock the
general-purpose processing power of RNICs and provide an unprecedented
level of programmability for complex offloads, by using novel
combinations of existing RDMA verbs (\S\ref{sec:design}).

\section{The \sys Computational Framework}
\label{sec:design}
To achieve our aforementioned goals, we develop a framework that enables complex offloads, called \sys.
\Sys's key idea is to combine widely available capabilities of RNICs
to enable self-modifying RDMA programs. These
programs---chains of RDMA operations---are capable of executing
dynamic control flows with conditionals and loops.
Fig.~\ref{fig:turing_complete} illustrates the usage of \sys. The setup phase involves (\circledgray{1}) preparing/compiling the RDMA code required for the service and (\circledgray{2}) posting the output chain(s) of RDMA WRs to the RNIC. Clients can then use the offload by invoking a trigger (\circledblue{3}) that causes the server's RNIC to (\circledgray{4}) execute the posted RDMA program, which returns a response (\circledblue{5}) to the client upon completion.

To further understand this proposed framework, we first look into the execution models offered by RNICs, and the ordering guarantees they provide for RDMA verbs. We then look into the expressivity of traditional RDMA verbs and explore parallels with CPU instruction sets. We use these insights to describe strategies for expressing complex logic using traditional RDMA verbs, \emph{without requiring any hardware modifications.}

\subsection{RDMA execution model}
\label{sec:consistency}

The RDMA interface specifies a number of data movement verbs
({\rread}, {\rwrite}, {\rsend}, {\rrecv}, etc.) that are \emph{posted}
as \emph{work requests} (WRs) by offload developers into \emph{work
  queues} (WQs) in host memory. The RNIC starts execution of a
sequence of WRs in a WQ once the offload developer triggers a
\emph{doorbell}---a special register in RNIC memory that informs the
RNIC that a WQ has been updated and should be executed.

\begin{figure}[t]
    \centering
    \qquad
    \subfloat[Completion order.\label{fig:completion_order}]{{\includegraphics[width=\columnwidth]{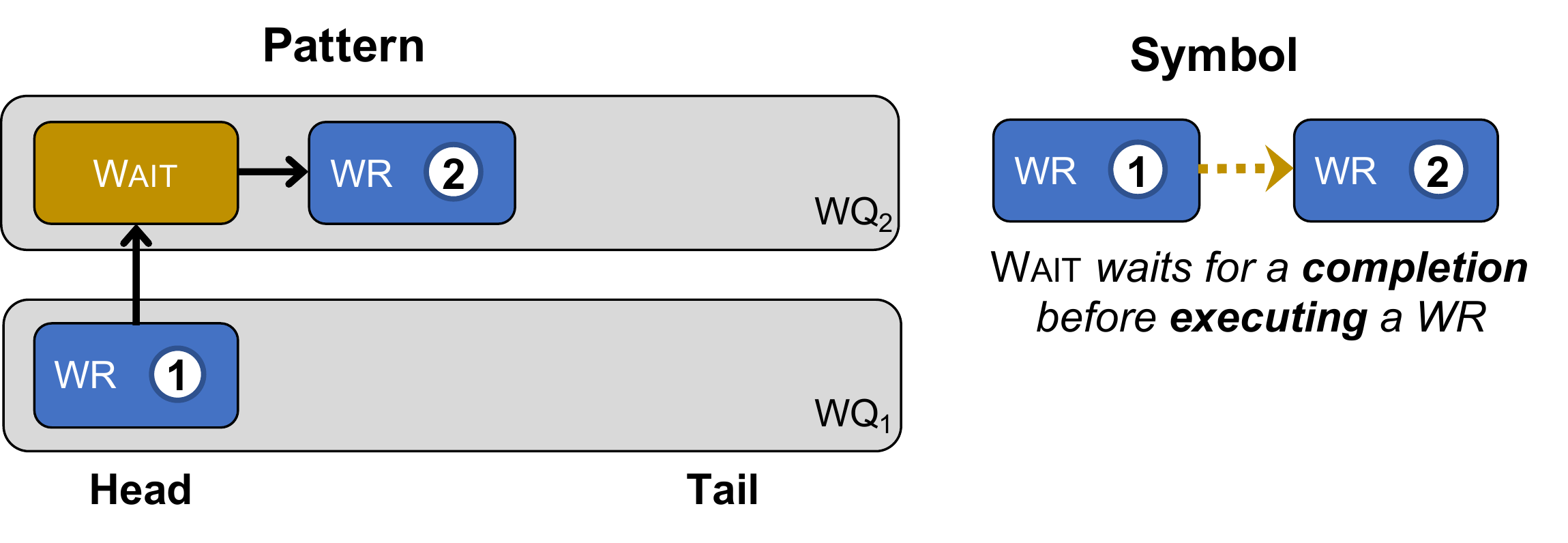}}}
    \qquad
    \subfloat[Doorbell order.\label{fig:doorbell_order}
    ]{{\includegraphics[width=\columnwidth]{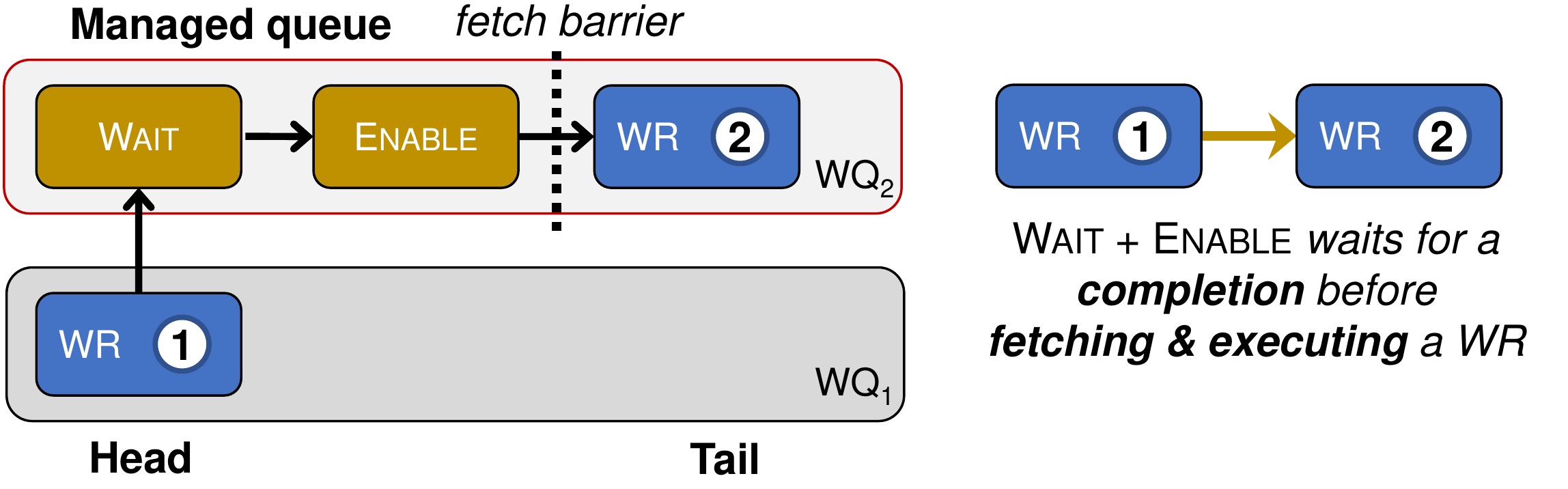}}}
    \caption{Work request ordering modes that guarantee a total order
      of operations \ref{fig:completion_order} and, a more restrictive
      ``doorbell'' order \ref{fig:doorbell_order}, where operations are fetched by the NIC one-by-one.
      The symbols on the right will be used as notation for these WR chains in the examples of \S\ref{sec:design}.}
    \label{fig:exec_model}
\end{figure}

\smartparagraph{Work request ordering.}
Ordering rules for RDMA WRs distinguish between write WRs and non-write WRs that return a value.
Within each category of operations, RDMA guarantees in-order execution of WRs within a single WQ.
In particular, write WRs (i.e., \rsend, \rwrite, \rwriteimm) are
totally ordered with regard to each other, but writes may be reordered before prior non-write WRs.%

We call the default RDMA ordering mode \emph{work queue (WQ) ordering}. Sophisticated offload logic often requires stronger ordering constraints, which we construct with the help of two RDMA verbs.
Fig.~\ref{fig:exec_model} shows two stricter ordering modes that we introduce and how to achieve them.

The {\rwait} verb stops WR execution until the completion
of a specified WR from another WQ or the preceding WR in the same WQ.
We call this \emph{completion ordering} (Fig.~\ref{fig:completion_order}). It achieves total ordering of WRs along the execution chain (which potentially involves multiple WQs).
It can be used to enforce data consistency, similar to data memory barriers in CPU
instruction sets---to wait for data to be available before executing the
WRs operating on the data.
Moreover, {\rwait} allows developers to \emph{pre-post} chains of RDMA verbs to the RNIC, without immediately executing them.

In all the aforementioned ordering modes, the RNIC is free to prefetch into its cache the WRs within a WQ.
Thus, the execution outcome reflects the WRs at the time they were
fetched, which can be incoherent with the versions that reside in host
memory in case these were later modified.  To avoid this issue, the
RNIC allows placing a WQ into \emph{managed} mode, in which WR
prefetch is disabled. The {\renable} verb is then used to explicitly start
the prefetching of WRs. This allows for 
existing WRs to be modified within the WQ, %
as long
as this is done before completion of the posted
{\renable}---similar to an instruction barrier.
We achieve
a full (data and instruction) barrier, by using {\rwait} and
{\renable} in sequence. We call this \emph{doorbell ordering}
(Fig.~\ref{fig:doorbell_order}). Doorbell ordering allows
developers to modify WR chains in-place. In particular, it allows for
\emph{data-dependent, self-modifying} WRs.

Thus, we have shown that we can control WR fetch and execution via special verbs, which
we will exploit in the next section to develop full-fledged RDMA programs.
These verbs are widely available in commodity RNICs
(e.g., Mellanox terms them cross-channel communication~\cite{mellanox-dev-guide}).

\subsection{Dynamic RDMA Programs}
\label{sec:dynamic}

\begin{figure}[t]
    \centering
    \includegraphics[width=\columnwidth]{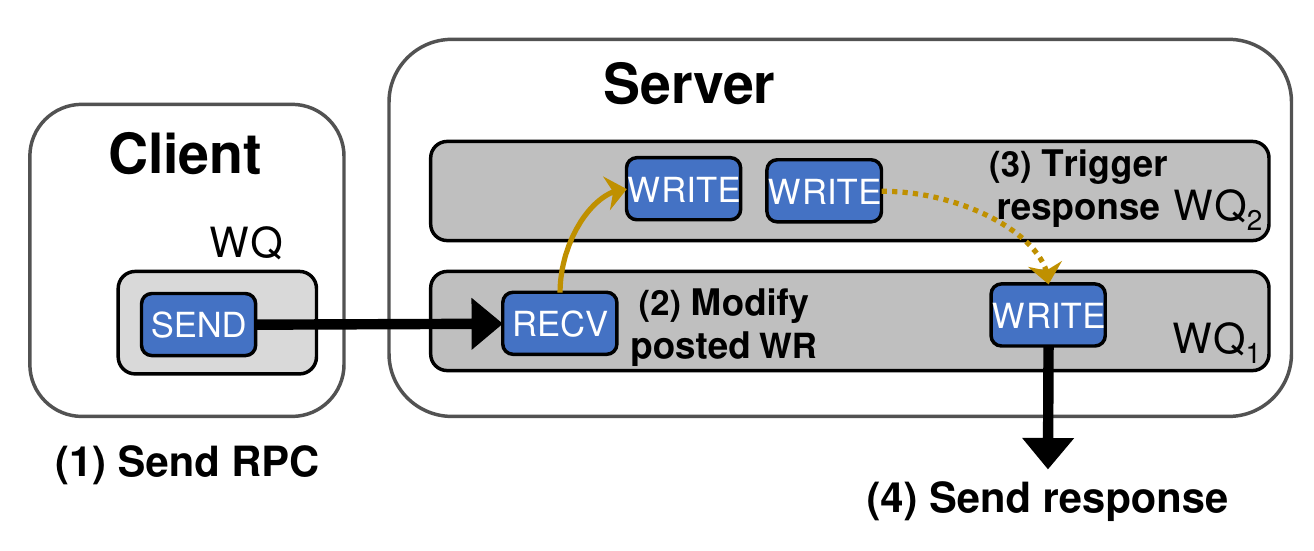}
    \caption{Clients can trigger posted operations. Thick solid lines represent (meta)data movements.}
    \label{fig:self_modifying}
\end{figure}

While a static sequence of RDMA WRs is already a rudimentary RDMA
program, complex offloads require \emph{data-dependent execution},
where the logic of the offload is dependent on input arguments. To
realize data-dependent execution, we construct \emph{self-modifying
  RDMA code}.
\paragraph{Self-modifying RDMA code.} Doorbell ordering enables a restricted form of self-modifying code, capable of
data-dependent execution.  To illustrate this concept, we use the
example of a server host that offloads an RPC handler to its RNIC as shown
in Fig.~\ref{fig:self_modifying}. The RPC response
depends on the argument set by the client and thus the RDMA offload
is data-dependent.  The server posts the RDMA program that
consists of a set of WRs spanning two WQs.  The client invokes the
offload by issuing a {\rsend} operation.  At the RNIC, the {\rsend}
triggers the posted {\rrecv} operation.
Observe that {\rrecv} specifies where the {\rsend} data is placed.  We
configure {\rrecv} to inject the received data into the posted WR
chain in $\mbox{WQ}_2$ to modify its attributes. We achieve this by leveraging
doorbell ordering, to ensure that posted WRs are not
prefetched by the RNIC and can be altered by preceding WRs.

This is an instance of self-modifying code. As such, clients
can pass arguments to the offloaded RPC handler and the RNIC will
dynamically alter the executed code accordingly.
However, this by itself is not sufficient to provide a Turing complete offload framework.

\paragraph{Turing completeness of RDMA.} Turing completeness implies
that a system of data-manipulation rules, such as RDMA, are
computationally universal. For RDMA to be Turing complete, we need to satisfy two requirements~\cite{gabbrielli2010programming}:\\
\textbf{T1:} Ability to read/write arbitrary amounts of memory.\\
\textbf{T2:} Conditional branching (e.g. if/else statements).

T1 can be satisfied for limited amounts of memory with regular RDMA verbs, whereas T2 has not been demonstrated with RDMA NICs.
However, to truly be capable of accessing an \emph{arbitrary} amount of memory, we need a way of realizing loops. Loops open up a
range of sophisticated use-cases and lower the number of constraints that programmers have to consider for offloads. To highlight their
importance, we add them as a third requirement, necessary to fulfill the first:\\
\textbf{T3:} The ability to execute code repeatedly (loops).

In the next sub-sections, we show how dynamic execution can be used to satisfy all the aforementioned requirements.
A proof sketch of Turing completeness is given in Appendix~\ref{appendix:turing}.

\subsection{Conditionals}
\label{sec:conditional}

\begin{figure}
    \centering
    \includegraphics[width=\columnwidth]{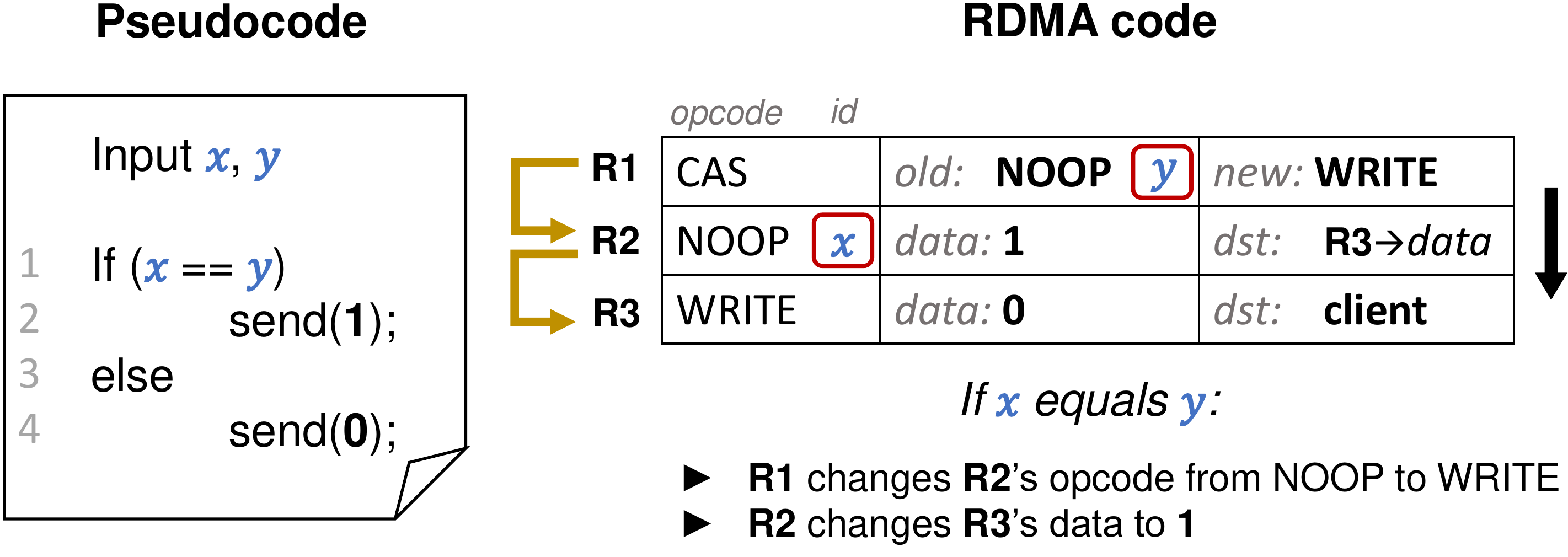}
    \caption{Simple \emph{if} example and equivalent RDMA code. Conditional execution
    relies on self-modifying code using CAS to enable/disable WRs based on the operand values.}
    \label{fig:if_example}
\end{figure}

Conditional execution---choosing what computation to perform based on
a runtime condition---is typically realized using conditional
branches, which are not readily available in RDMA.  To this end, we
introduce a novel approach that uses self-modifying {\rcas}
verbs.  The main insight is that this verb can be used to check a
condition (\ie equality of $x$ and $y$) and then perform a swap to
modify the attributes of a WR.  We describe how this is done in
Fig.~\ref{fig:if_example}. We insert a {\rcas} that compares the
64-bit value at the address of R2's \emph{opcode} attribute (initially
{\rnoop}) with its \emph{old} parameter (also initially {\rnoop}). We
then set the \emph{id} field of R2 to $x$. This field can be
manipulated freely without changing the behavior of the WR, allowing
us to use it to store $x$. Operand $y$ is stored in
the corresponding position in the \emph{old} field of R1. This means
that if $x$ and $y$ are equal, the {\rcas} operation will succeed and
the value in R1's \emph{new} field---which we set to {\rwrite}---will
replace R2's opcode. Hence, in the case $x=y$, R2 will change from a
{\rnoop} into a {\rwrite} operation. This {\rwrite} is set to modify
the \emph{data} value of the return operation (R3) to 1. If $x$ and
$y$ are not equal, the default value 0 is returned.

Now that we have established the utility of this technique for basic conditionals, we next look
into how to can be used to support loop constructs.

\subsection{Loops}
\label{sec:loops}

To support loop constructs efficiently, we require (1) conditional branching to test the loop condition and break if necessary, and (2) WR re-execution, to repeat the loop body. We develop each, in turn, below.

\begin{figure}
    \centering
    \includegraphics[width=\columnwidth]{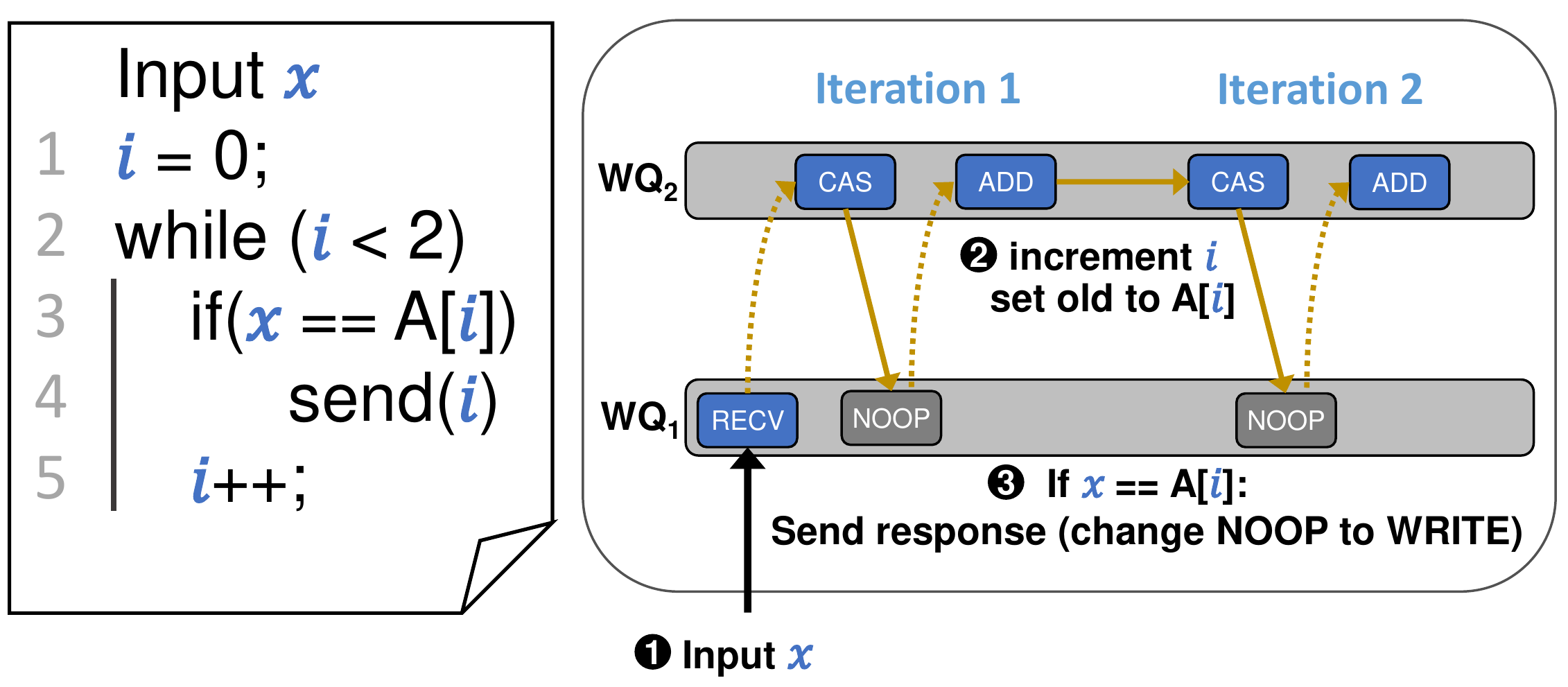}
    \caption{\textsf{while} loop using {\rcas}. Loop is unrolled since loop size is fixed and set to 2.}
    \label{fig:while_example_simple}
\end{figure}

Consider the while loop example in
Figure~\ref{fig:while_example_simple}.  This offload searches for $x$
in an array $A$ and sends the corresponding index. The loop is
static because $A$ has finite size (in this case, size =2), known a
priori.  To simplify presentation, consider the case
$A[i] = i, \forall i$.  Without this simplification, the example would
include an additional {\rwrite} to fetch the value at $A[i]$.

The loop body uses a {\rcas} verb to
implement the if condition (line 3), followed by an {\radd} verb to increment
$i$ (line 6). Given that the loop size is known a priori ($\mbox{size}=2$), \sys can unroll the while
loop in advance and post the WRs for all iterations. As such, there is no need to check
the condition at line 2. For each iteration, if the {\rcas} succeeds, the {\rnoop} verb
in $WQ_{1}$ will be changed to {\rwrite}---which will send the response back to the client.
However, it is clear that, regardless of the comparison result, all subsequent iterations will
be executed. This is inefficient since, if the send (line 4) occurs before the loop is finished,
a number of WRs will be wastefully executed by the NIC. This is impractical
for larger loop sizes or if the number of iterations is not known a priori.

\begin{figure}
    \centering
    \includegraphics[width=\columnwidth]{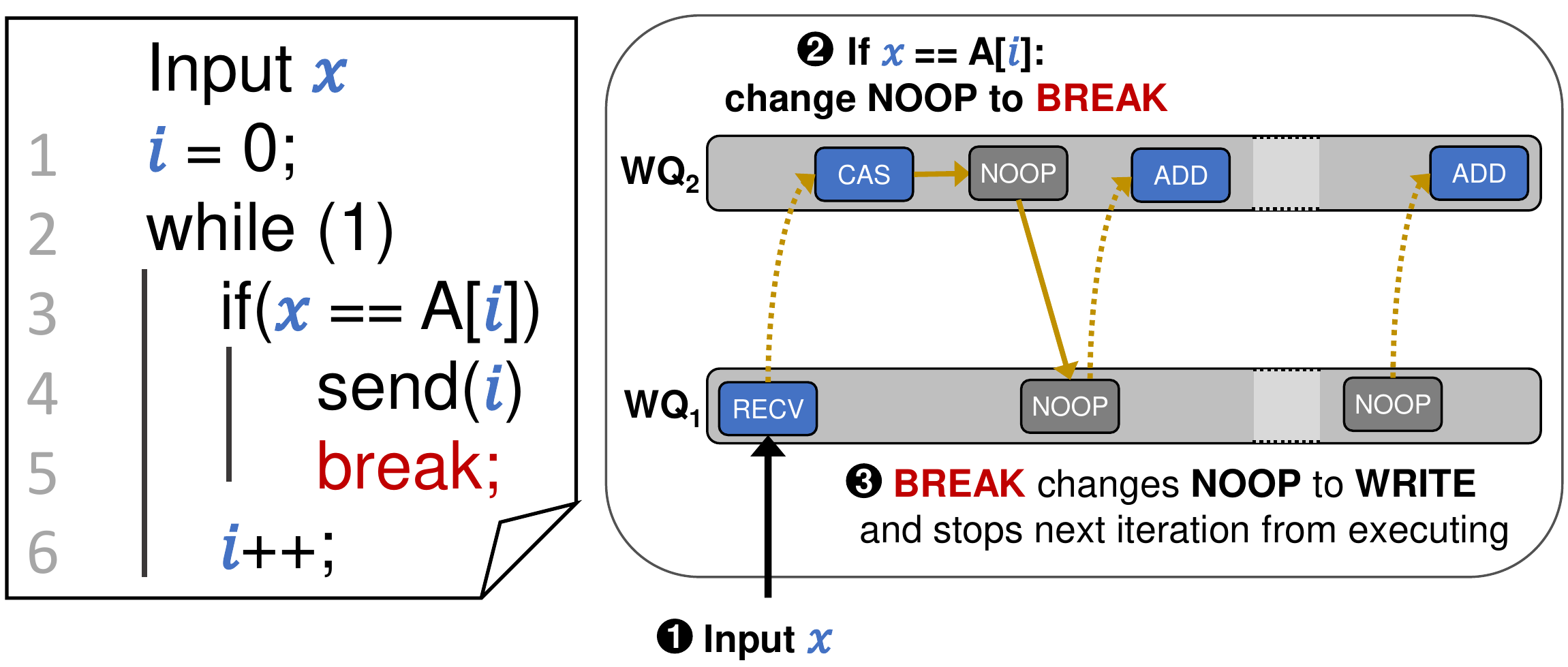}
    \caption{\emph{while} loop with breaks realized using CAS. 
    To implement breaks, we use {\rcas} to change a {\rnoop} WR to an RDMA {\rwrite}, which then stops subsequent iterations from executing.}
    \label{fig:while_example_break}
\end{figure}

\paragraph{Unbounded loops and termination.}
\label{sec:breaks}
Figure~\ref{fig:while_example_break} modifies the previous example
to make it such that the loop is unbounded. For efficiency,
we add a \textsf{break} that exits the loop if the element is found.%
The role of \textsf{break} is to prevent additional iterations from being executed.  We use an additional 
 {\rnoop} that is formatted such that, once transformed into a
{\rwrite} by the {\rcas} operation, it prevents the execution of subsequent iterations in the
loop. This is done by modifying the last WR in the loop such that it
does not trigger a completion event. The next iteration in the loop,
which {\rwait}s on such an event (via completion ordering), will therefore not be executed. Moreover, the {\rwrite}
will also modify the opcode of the WR used to send back the response from {\rnoop} to {\rwrite}.

As such, \textsf{break} allows efficient and unbounded loop execution. However, it still remains necessary for the CPU
to post WRs to continue the loop after all its WRs are executed. This consumes CPU cycles and can even
increase latency if the CPU is unable to keep up with the speed of WR execution. %

\paragraph{Unbounded loops via WQ recycling.}
\label{sec:recycling}
To allow the NIC to recycle WRs without CPU intervention, we make use of a novel
technique that we call \emph{WQ recycling}.
RNICs iterate over WQs, which are circular
buffers, and execute the WRs therein.  By design, each WR is
meant to execute only once. %
However, there is no fundamental reason why WRs cannot be reused since the
RNIC does not actually erase them from the WQ. To enable recycling of a WR
chain, we insert a {\rwait} and {\renable} sequence at the tail of the WQ. This
instructs the RNIC to wrap around the tail and re-execute the WR chain for as many
times as needed.

It is important to note that WQ recycling is not a panacea. To allow the tail of the
WQ to wrap around, all posted {\rwait} and {\renable} WRs in the loop need to have their \emph{wqe\_count}
attribute updated. This attribute is used to determine the index of the WR that these ordering verbs affect. In ConnectX NICs,
these indices are maintained internally by the RNIC and their values are monotonically increasing (instead of resetting after the WQ wraps
around). As such, the \emph{wqe\_count} values need to be incremented to match.
This incurs overhead (as seen in Table
\ref{table:constructs}) and requires an additional {\radd} operation in combination with other verbs.
As such, loop unrolling, where each iteration is manually posted by the CPU, is overall less taxing on the RNIC.
However, WQ recycling avoids CPU intervention, allowing the offload to remain available even amid host
software failures (as we will see later in \S\ref{sec:recovery}).

\subsection{Putting it all together} %
With conditional branching, we can dynamically alter the control flow
of any function on an RNIC. Loops allow us to traverse arbitrary data structures. Together, we
have transformed an RNIC into a general processing unit. In this
section, we discuss the usability aspects from overhead, security,
programmability, and expressiveness perspectives.

\smartparagraph{Building blocks.} We abstract and parameterize the RDMA chains
required for conditional branching and looping into \textsf{if} and
\textsf{while} constructs. The overhead in terms of RDMA WR chains
of our constructs is shown in Table~\ref{table:constructs}. We can see a
breakdown of the minimum number of operations
required for each.
Inequality predicates, such as $<$ or $>$, can also be supported by
combining equality checks with {\rmax} or {\rmin}, as seen later in
Table~\ref{table:verbs}. However, their availability is
vendor-specific and currently only supported by ConnectX NICs.

\begin{table}[t]
\resizebox{\columnwidth}{!}{%
\begin{tabular}{|c|c|c|c|}
\hline
\multicolumn{2}{|c|}{\textbf{\sys Constructs}} & \textbf{Number of WRs} & \textbf{Operand limit {[}bits{]}} \\ \hline
\multicolumn{2}{|c|}{if}              & 1C + 1A + 3E   & \multirow{3}{*}{48}      \\ \cline{1-3}
\multirow{2}{*}{while}   & Unrolled   & 1C + 1A + 3E   &                          \\ \cline{2-3}
                         & Recycled   & 3C + 2A + 4E   &                          \\ \hline
\end{tabular}

}
\caption{Breakdown of the overhead of our constructs with different offload strategies.
\emph{C} refers to copy verbs, \emph{A} refers to atomic RDMA verbs, and \emph{E} refers to {\rwait}/{\renable} verbs.
\textsf{while} loops that use WQ recycling incur 2 additional {\rread}s, 1 {\radd}, and 1 {\renable} WR.}
\label{table:constructs}
\end{table}

\smartparagraph{Operand limits.}
\sys's limit is based on the
supported size for the {\rcas} verb, which is 64 bits. The operand
is provided as a 48-bit value, encoded in its \emph{id} and other
neighboring fields (which can also be freely modified without
affecting execution). The remaining bits are used for modifying the opcode of the WR
depending on the result of the comparison. We note that our advertised limits only signify what is possible
with the number of operations we allocate for our constructs. For
instance, despite the 48-bit operand limit for our constructs,
we can chain together multiple CAS operations to handle different
segments of a larger operand (we do not rely on the atomicity property of CAS). As such, there is no fundamental
limitation, only a performance penalty.

\smartparagraph{Offload setup.} To offload an RDMA program, clients
first create an RDMA connection to the target server and send an RPC
to initiate the offload. We envision that the server already has the
offload code; however, other ways of deploying the offload are
possible.  Upon receiving a connection request, the server creates one
or more managed local WQs to post the offloaded code. Next, it
registers two main types of memory regions for RDMA access: (a) a code
region, and (b) a data region.  The code region is the set of remote
RDMA WQs created on the server, which are unique to each client and
need to be accessible via RDMA to allow self-modifying code. Code
regions are protected by memory keys---special tokens required for
RDMA access---upon registration (at connection time), prohibiting
unauthorized access. The data region holds any data elements used by
the offload (\eg a hash table). Data regions can be shared or private,
depending on the use-case.

\smartparagraph{Security.}
\sys does not solve security challenges in existing RDMA or Infiniband
implementations~\cite{simpson2020securing}. However, \sys can help
RDMA systems become more secure. %
For such systems, \emph{one-sided} RDMA operations (\eg RDMA \rread
and \rwrite) are frequently
used~\cite{farm,pilaf,drtm,octopus,hyperloop,drtmh} as they avoid CPU
overheads at the responder. However, doing so requires clients to have
direct read and/or write memory access. This can compromise security
if clients are buggy and/or malicious. To give an example, FaRM allows
clients to write messages directly to shared RPC buffers. This
requires clients to behave correctly, as they could otherwise
overwrite or modify other clients' RPCs.  \Sys allows applications to
use \emph{two-sided} RDMA operations (\eg \rsend and \rrecv), which do
not require direct memory access, while \emph{still} fully bypassing
server CPUs. As we demonstrate in our use-cases in \S\ref{sec:eval},
\rsend operations can be used to trigger offload programs without any
CPU involvement.

\smartparagraph{Isolation.} Given that \sys implements dynamic loops, clients can abuse
such constructs to consume more than their fair share of resources. Luckily, popular RNICs, like ConnectX, provide WQ rate-limiters~\cite{ratelimit} for performance isolation. As such, even if clients trigger
non-terminating offload code, they still have to adhere to their assigned rates. Moreover, offloaded code can be configured by the servers to be auditable through completion events, created automatically after a WR is executed.
These events can be monitored and servers can terminate connections to clients running misbehaving code.

\paragraph{Parallelism.} RDMA WR fetch and execution latencies are more costly
compared to CPU instructions, as WRs are fetched/executed via PCIe (microseconds vs.
nanoseconds). As such, to hide WR latencies, it is important to parallelize
logically unrelated operations. Like threads of execution in a CPU, each WQ is
allocated a single RNIC PU to ensure in-order execution without inter-PU
synchronization. As such, we carefully tune our offloaded code to allow unrelated verbs to
execute on independent queues to be able to
parallelize execution as much as possible. The benefits of parallelism are
evaluated in \S\ref{sec:hash}.

\section{Implementation}
\label{sec:impl}

Our offload framework is implemented in C with $\sim$2,300 lines of code---this includes our use cases ($\sim$1400), and convenience wrappers for RDMA verbs (\emph{libibverbs}) API ($\sim$900).

Our approach does not require modifying any RDMA
libraries or drivers. \sys uses low-level functions provided
by Mellanox's ConnectX driver (\emph{libmlx5}) to expose
in-memory WQ buffers and register them to the RNIC,
allowing WRs to be manipulated via RDMA verbs.
We configure the ConnectX-5 firmware to allow the WR \emph{id} field
to be manipulated freely, which is required for conditional operations as well as WR recycling.
This is done by modifying specific configuration registers on the NIC~\cite{pcx}.

\sys is compatible with any ConnectX NICs that
support \rwait and \renable (\eg ConnectX-3 and later models).

\section{Evaluation}
\label{sec:eval}
We start by characterizing the underlying RNIC performance (\S\ref{sec:micro}) to understand how it affects our implemented programming constructs.
Then, in our evaluation against state-of-the-art RNIC and SmartNIC offloads, we show that \sys:

\begin{enumerate}[noitemsep, nolistsep]
\item Speeds up remote data structure traversals, such as hash tables
  (\S\ref{sec:hash}) and linked lists (\S\ref{sec:linkedlist})
  compared to vanilla RDMA offload;
  
\item Accelerates (\S\ref{sec:memcached}) and provides performance
  isolation (\S\ref{sec:perf_isolation}) for the Memcached key-value
  store;
  
\item Provides improved availability for applications (\S\ref{sec:recovery})---allowing
  them to run in spite of OS \& process crashes;
  
\item Exposes programming constructs generic enough to enable a
  wide-variety of use-cases (\S\ref{sec:hash}--\S\ref{sec:recovery});
\end{enumerate}

\smartparagraph{Testbed.} Our experimental testbed consists of 3$\times$ dual-socket Haswell servers running at 3.2 GHz, with a total of 16 cores, 128 GB of DRAM, and 100 Gbps dual-port Mellanox ConnectX-5 Infiniband RNICs. All nodes are running Ubuntu 18.04 with Linux Kernel version 4.15 and are connected via back-to-back Infiniband links. %

\smartparagraph{NIC setup.} For all of our experiments, we use reliable connection (RC) RDMA transport, which supports the RDMA synchronization features we use. All WQs that enforce doorbell order are initialized with a special ``managed'' flag to disable the driver from issuing doorbells after a WR is posted. The WQ size is set to match that of the offloaded program.

\subsection{Microbenchmarks}
\label{sec:micro}

We run microbenchmarks to break down RNIC verb execution latency,
understand the overheads of our different ordering modes, and
determine the processing bandwidth of different RDMA verbs and of our
constructs.

\begin{figure}[t]
    \centering
    \includegraphics[width=\columnwidth]{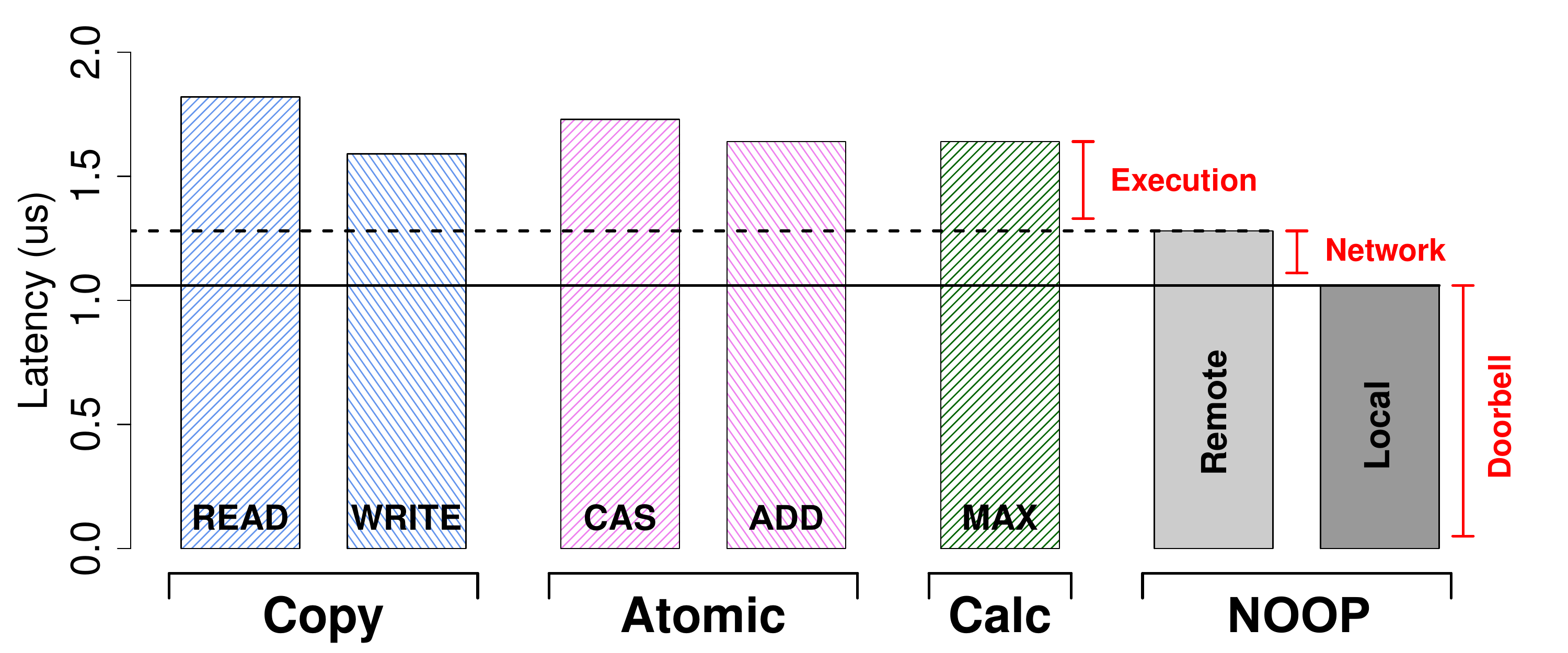}
    \caption{Latencies of different RDMA verbs. The solid line marks the latency of ringing the doorbell via MMIO. The difference between dashed and solid lines estimates network latency.}
    \label{fig:verb_lat}
\end{figure}

\subsubsection{RDMA Latency}
We break down the performance of RDMA verbs, configured to perform 64B
IO, by measuring their average latencies after executing them 100K
times. All verbs are executed remotely, unless otherwise stated. As seen
in Fig.~\ref{fig:verb_lat}, {\rwrite} has a latency of 1.6 \us. It
uses posted PCIe transactions, which are one-way.  Comparatively,
non-posted verbs such as {\rread} or atomics such as fetch-and-add
({\radd}) and compare-and-swap ({\rcas}) need to wait for a PCIe
completion and take $\sim$1.8 \us.\footnote{Older-generation NICs (\eg
  ConnectX-4) use a proprietary concurrency control mechanism to
  implement atomics, resulting in higher latencies than later
  generations that rely on PCIe atomic transactions.~} Overall, the execution time difference is small among
verbs, even for more advanced, vendor-specific \emph{Calc} verbs that
perform logical and arithmetic computations (\eg{} {\rmax}).

To break down the different latency components for RDMA verb execution,
we first estimate the latency of issuing a doorbell and copying
the WR to the RNIC. This can be done by measuring the execution
time of a {\rnoop} WR. This time can be subtracted from the
latencies of other WRs to give an estimate of their execution
time once the WR is available in the RNIC's cache. We also
quantify the network cost by executing remote and local loopback
{\rnoop} WRs (shown on the right-hand side) and measuring
the difference---roughly 0.25 \us for our back-to-back connected nodes.
Overall, these results show low verb execution latency, justifying building more sophisticated functions atop. We next measure the implications of ordering for offloads.

\subsubsection{Ordering Overheads}
\label{sec:overheads}
\begin{figure}[t]
    \centering
    \includegraphics[width=0.5\textwidth]{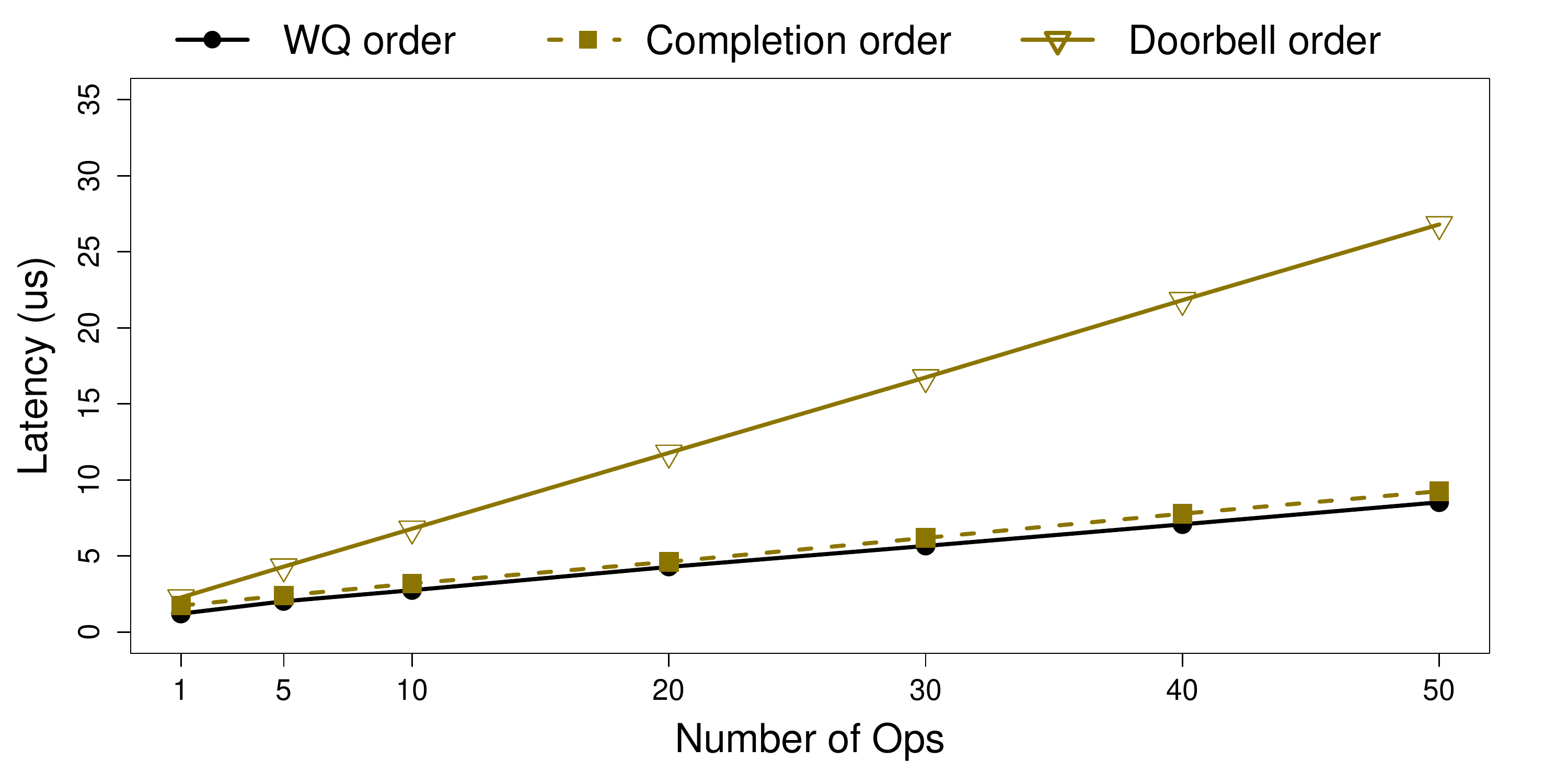}
    \caption{Execution latency of RDMA verbs posted using different ordering modes. More restrictive modes such as Doorbell order add non-negligible overheads as it requires the NIC to fetch WRs sequentially.}
    \label{fig:group_lat}
\end{figure}

We show the latency of executing chains of RDMA verbs using different
ordering modes. All the posted WRs within a chain are {\rnoop}, to
simplify isolating the performance impact of ordering. We start by
measuring the latency of executing a chain of verbs posted to the same
queue but absent any constraints (WQ order), and compare it to the
ordering modes that we introduced in
Fig.~\ref{fig:exec_model}---completion order and doorbell order.  WQ
order only mandates in-order updates to memory, which allows for
increased concurrency. Operations that are not modifying the same
memory address can execute concurrently and the RNIC is free to
prefetch multiple WRs with a single DMA\footnote{The number of
  operations fetched by the RNIC can change dynamically. The Prefetch
  mechanism in ConnectX RNICs is proprietary.}.
We can see in Fig.~\ref{fig:group_lat} that the latency of a single {\rnoop} is 1.21 \us
and the overhead of adding subsequent verbs is roughly
0.17 \us per verb. The first verb is slower since it requires an
initial doorbell to tell the NIC that there is outstanding work. For
completion ordering, less concurrency is possible since WRs await the
completions of their predecessors, and the overhead of increases slightly
to 0.19 \us per additional WR.
For doorbell order, no latency-hiding is possible, as the NIC has
to fetch WRs from memory one-by-one, which results in an overhead
of 0.54 \us per additional WR. These results signify that, doorbell
ordering should be used conservatively, as there is more than 0.5
\us latency increase for every instance of its use, compared to
more relaxed ordering modes.

\subsubsection{RDMA Verb Throughput}

We show the throughput of the common RDMA verbs in Table~\ref{table:verbs} for a single ConnectX-5 port.
ConnectX cards assign compute resources on a per port basis. For ConnectX-5, each port has 8 PUs.
Atomic verbs, such as CAS, offer a
comparatively limited throughput (8$\times$ lower than regular verbs)
due to memory synchronization across PCIe.

In addition, we measure the performance of \sys's \textsf{if} and \textsf{while} constructs.
Using 48-bit operands, a ConnectX-5 NIC can execute 700K \textsf{if}
constructs per second. This is due to the need for CAS to ensure doorbell ordering between CAS and the subsequent WR it modifies. This causes the throughput to be bound by NIC processing limits. Unrolled \textsf{while} loops require the same number of verbs per iteration as an \textsf{if} statement and their throughput is identical. \textsf{while} loops with WQ recycling have reduced performance due to having to execute more WRs per iteration. %

\begin{table}
\resizebox{\columnwidth}{!}{%
\begin{tabular}{|c|c|c|c|c|}
\hline
\multicolumn{3}{|c|}{\multirow{2}{*}{\textbf{Operation}}}             & \multirow{2}{*}{\textbf{Throughput (M ops/s)}} & \multirow{2}{*}{\textbf{Support}}             \\
\multicolumn{3}{|c|}{}   &                                       &                                      \\ \hline
\multirow{2}{*}{Atomic}  & \multicolumn{2}{c|}{CAS}          & \multirow{2}{*}{8.4}                  & \multirow{4}{*}{Native}              \\ \cline{2-3}
                         & \multicolumn{2}{c|}{ADD}          &                                       &                                      \\ \cline{1-4}
\multirow{2}{*}{Copy}    & \multicolumn{2}{c|}{READ}         & 65                                    &                                      \\ \cline{2-4}
                         & \multicolumn{2}{c|}{WRITE}        & 63                                    &                                      \\ \hline
Calc                     & \multicolumn{2}{c|}{MAX}          & 63                                    & Mellanox                             \\ \hline
\multirow{3}{*}{Constructs} & \multicolumn{2}{c|}{if}           & 0.7                                   & \multirow{3}{*}{\sys} \\ \cline{2-4}
                         & \multirow{2}{*}{while} & Unrolled & 0.7                                   &                                      \\ \cline{3-4}
                         &                        & Recycled & 0.3                                   &                                      \\ \hline
\end{tabular}
}
\caption{Throughput of common RDMA verbs and \sys's constructs on a single port of a ConnectX-5. \textsf{if} and unrolled \textsf{while} have identical performance. \textsf{while} loops with WQ recycling require additional WRs and therefore have a lower throughput.
}
\label{table:verbs}
\end{table}

\subsection{Offload: Hash Lookup}
\label{sec:hash}

After evaluating the overheads of \sys's ordering modes and
constructs, we next look into the performance of \sys for offloading
remote access to popular data structures. We first look into hash
tables, given their prominent use in key-value stores for indexing stored objects. To perform a simple \emph{get} operation, clients first have to lookup the desired key-value entry in the hash table. The entry can either have the value directly inlined or a pointer to its memory address. The value is then fetched and returned back to the client. Hopscotch hashing is a popular hashing scheme that resolves collisions by using \emph{H} hashes for each entry and storing them in 1 out of \emph{H} buckets. Each bucket has a \emph{neighborhood} that can probabilistically hold a given key. A lookup might require searching more than one bucket before the matching key-value entry is found. To support dynamic value sizes, we assume the value is not inlined in the bucket and is instead referenced via a pointer.

For distributed key-value stores built with RDMA, \emph{get} operations are usually implemented in one of two ways:

\smartparagraph{One-sided} approaches first retrieve the key's location using a one-sided RDMA READ operation and then issue a second READ to fetch the value. These approaches typically require two network round-trips at a minimum. This greatly increases latency but does not require involvement of the server's CPU. Many systems utilize this approach to implement lookups, including FaRM~\cite{farm} and Pilaf~\cite{pilaf}.

\smartparagraph{Two-sided} approaches require the client to send a
request using an RDMA SEND or WRITE. The server intercepts the
request, locates the value and then returns it using one of the
aforementioned verbs. This widely used~\cite{assise, herd} approach
follows traditional RPC implementations and avoids the need for
several roundtrips. However, this comes at the cost of server CPU
cycles.

\subsubsection{\sys's Approach}
To offload key-value \emph{get} operations, we leverage the offload
schemes introduced in \S\ref{sec:conditional} and \S\ref{sec:loops}.

\begin{figure}
    \centering
    \includegraphics[width=0.48\textwidth]{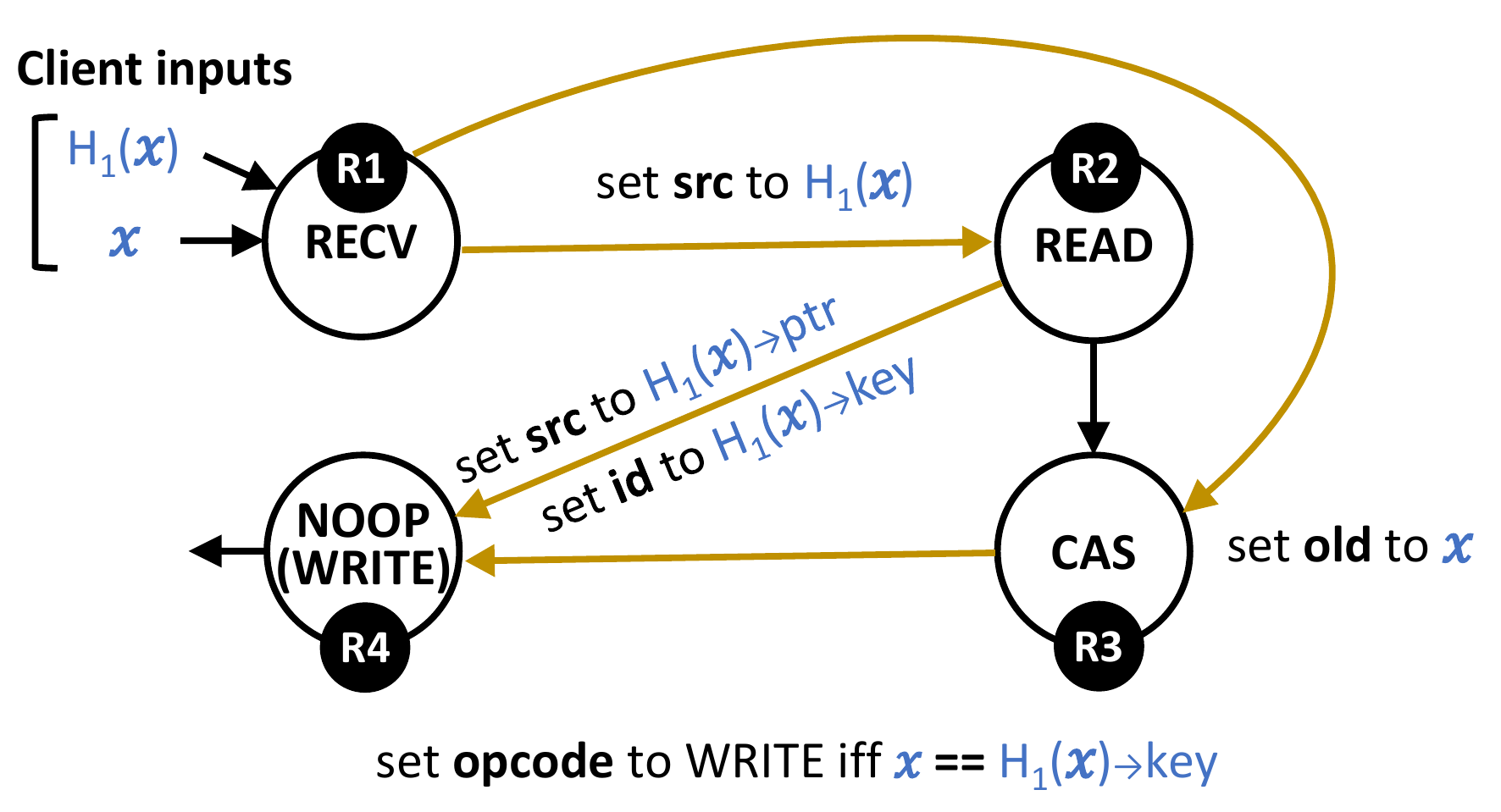}
    \caption{Hash lookup RDMA program. Black arrows indicate order of
      execution of WRs in their WQs. Brown arrows indicate
      self-modifying code dependencies and require doorbell
      ordering. $x$ is the requested key and $H_{1}(x)$ is its first
      hash. The acronym \emph{src} indicates the ``source address''
      field of WRs. \emph{old} indicates the ``expected value''
      at the target address of the {\rcas} operation. The \emph{id} field is used for storing conditional operands. %
    }
    \label{fig:hashget_overview}
\end{figure}

Fig.~\ref{fig:hashget_overview} describes the RDMA operations involved
for a single-hash lookup. To \emph{get} a value corresponding to a key,
the client first computes the hashes for its key. For this use-case, we set the
number of hashes to two, which is common in practice~\cite{memc3}. The
client then performs a \rsend with the value of the key $x$ and address of the first
bucket $H_{1}(x)$,
which are then captured via a \rrecv WR posted on the server. The
\rrecv WR (\circledblack{R1}) inserts $x$ into the $old$ field of the \rcas WR (\circledblack{R3}) and the bucket
address $H_{1}(x)$ into the \rread WR (\circledblack{R2}). The \rread WR retrieves the bucket and sets the source address (\emph{src})
of the response WR (\circledblack{R4}) to the address of the value (\emph{ptr}). It
also inserts the bucket's key into the \emph{id} field to prepare it for the conditional
check. Finally, \rcas (\circledblack{R3}) checks whether the expected value $old$, which is set
to key $x$, matches the \emph{id} field in (\circledblack{R4}), which is set to the
bucket's $key$. If equal, (\circledblack{R4})'s opcode is changed from
\rnoop to \rwrite, which then returns the value from the bucket. Given that each key may be stored in multiple buckets
(two in our setup), these lookups may be performed sequentially or in parallel, depending on
the offload configuration.

\subsubsection{Results}

\begin{figure}[b]
    \centering
    \includegraphics[width=0.5\textwidth]{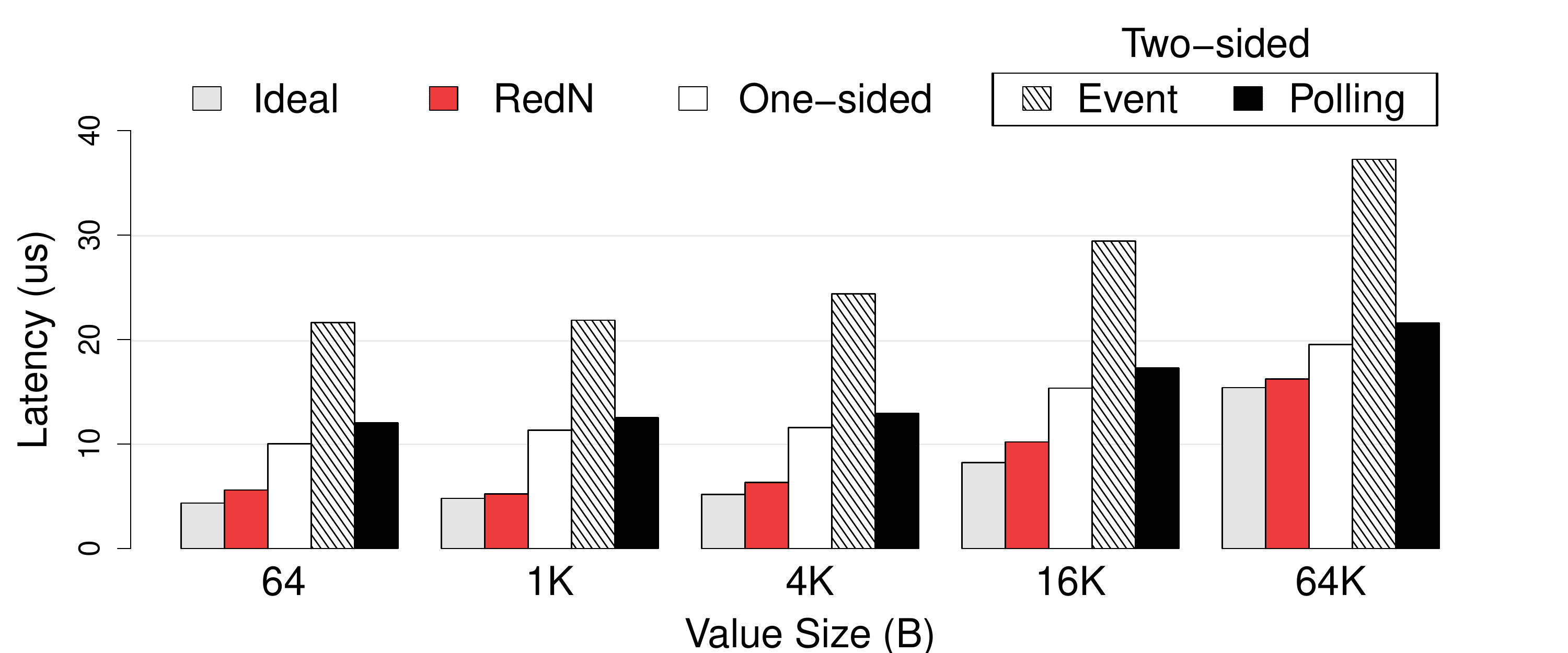}
    \caption{Average latency of hash lookups. \emph{Ideal} shows the latency of a single network round-trip \rread.
    }
    \label{fig:hash_lat}
\end{figure}

\begin{figure}[t]
    \centering
    \includegraphics[width=0.5\textwidth]{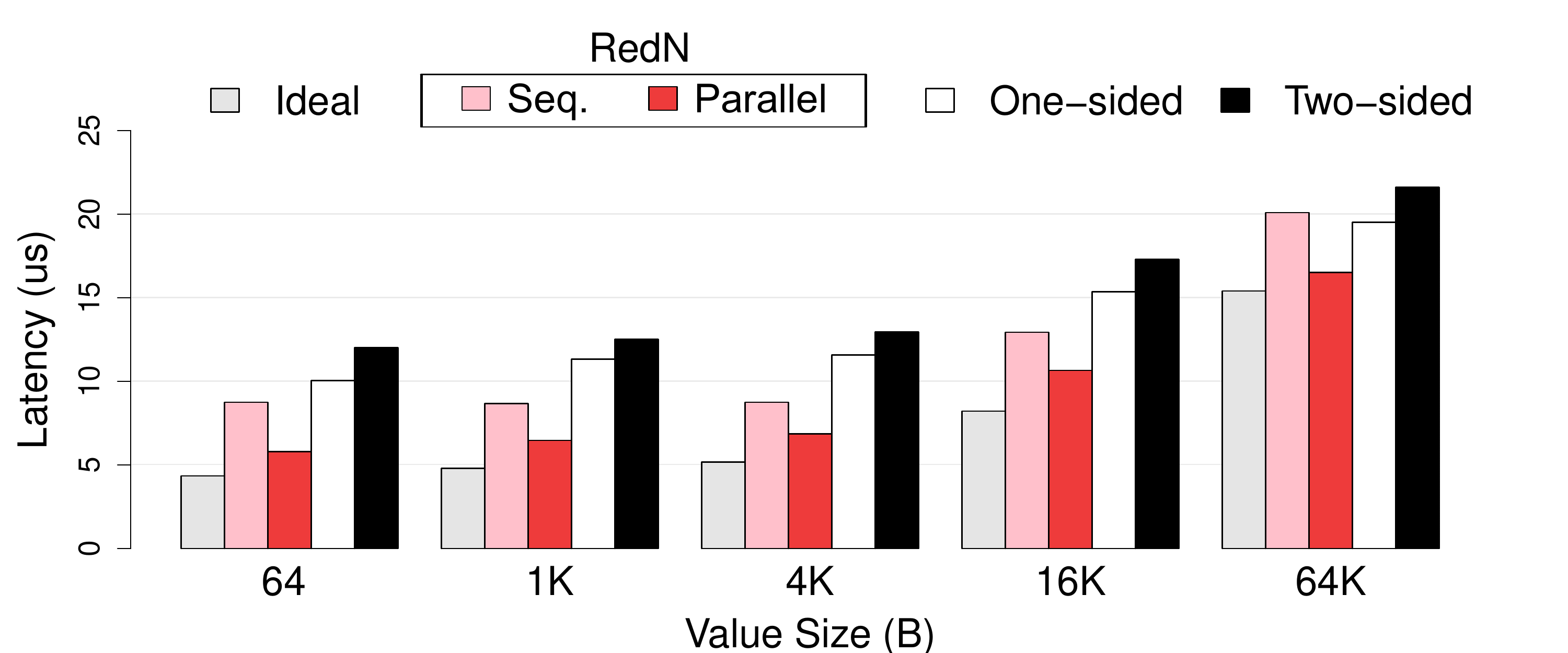}
    \caption{Average latency of hash lookups during collisions. \emph{Ideal} shows the latency of a single network round-trip \rread.}
    \label{fig:hash_collision_lat}
\end{figure}

We evaluate our approach against both one-sided and two-sided
implementations of key-value \emph{get} operations. We use FaRM's
approach~\cite{farm} to perform one-sided lookups. FaRM uses Hopscotch
hashing to locate the key using approximately two RDMA READs --- one
for fetching the buckets in a neighborhood that hold the key-value
pairs and another for reading the actual value. The neighborhood size
is set to 6 by default, implying a 6$\times$ overhead for RDMA
metadata operations. For two-sided lookups, our RPC to the
host involves a client-initiated RDMA {\rsend} to transmit the
\emph{get} request, and an RDMA {\rwrite} initiated by the server to
return the value after performing the lookup.

\smartparagraph{Latency.} Fig. \ref{fig:hash_lat} shows a latency
comparison of KV \emph{get} operations of \sys against one-sided and
two-sided baselines. We evaluate two distinct variations of two-sided.
The \emph{event-based} approach blocks for a completion event
to avoid wasting CPU cycles, whereas the \emph{polling-based} approach dedicates
one CPU core for polling the completion queue.
We use 48-bit keys and vary the value size. The
value size is given on the x-axis. In this
scenario, we assume no hash collisions and that all keys are found in
the first bucket. \sys is able to outperform all baselines ---
fetching a 64 KB key-value pair in 16.22 \us, which is within 5\% of a
single network round-trip \rread (Ideal). \sys is able to deliver
close-to-ideal performance because it bypasses the server's CPU
\textit{\textbf{and}} fetches the value in a single network
RTT. Compared to \sys, one-sided operations incur up to 2$\times$
higher latencies, as they require two RTTs to fetch a value.
Two-sided implementations do not incur any extra RTT; however, they require server CPU
intervention. The polling-based variant consumes an entire CPU core but 
provides competitive latencies. Event-based approaches block for completion events
to avoid wasting CPU cycles and incur much higher latencies as a consequence. \sys is able
to outperform polling-based and event-based approaches by up to 2 and 3.8$\times$,
respectively. Given the much higher latencies of event-based approaches, for the remainder
of this evaluation, we will only focus on polling-based approaches and simply refer to them
hereafter as \emph{two-sided}.

Fig. \ref{fig:hash_collision_lat} shows the latency in the presence of
hash collisions. In this case, we assume a worst case scenario, where
the key-value pair is always found in the second bucket. In this
scenario, we introduce two offload variants for \sys --- \sysseq \&
\syspar. The former performs bucket lookups sequentially within a
single WQ. The latter parallelizes bucket lookups by performing the
lookups across two different WQs to allow execution on different NIC
PUs. We can see that \syspar maintains similar latencies to lookups
with no hash collisions (\ie \emph{\sys} in Fig.~\ref{fig:hash_lat}),
since bucket lookups are almost completely parallelized. It is worth
noting that parallelism in this case does not cause unnecessary data
movement, since the value is only returned when the corresponding key
is found. For the other bucket, the \rwrite operation (R4 in
Fig.~\ref{fig:hashget_overview}) is a \rnoop. \sysseq, on the other
hand, incurs at least 3 \us of extra latency as it needs to search the
buckets one-by-one. As such, whenever possible, operations with no
dependencies should be executed in parallel. The trade-off is having
to allocate extra WQs for each level of parallelism.

\smartparagraph{Throughput.} We describe our throughput in Table~\ref{table:hash_tput}.
At lower IO, \sys is bottlenecked by the NIC's
processing capacity due to the use of doorbell ordering---reaching 500K ops/s on a single port (1M ops/s with dual ports).
At 64 KB, \sys reaches the single-port IB bandwidth limit (\textasciitilde~92 Gbps). Dual-port configs are limited by ConnectX-5's 16$\times$ PCIe 3.0 lanes.

\begin{table}[]
\centering
\resizebox{0.9\columnwidth}{!}{%
\begin{tabular}{l|c|c|c|c|}
\cline{2-5}
\multicolumn{1}{c|}{\multirow{2}{*}{\textbf{Hash lookup}}} & \multicolumn{4}{c|}{\textbf{IO Size}}                             \\ \cline{2-5} 
\multicolumn{1}{c|}{}                             & \multicolumn{2}{c|}{\textbf{$\leq$ 1 KB}}   & \multicolumn{2}{c|}{\textbf{64 KB}} \\ \hline
\multicolumn{1}{|l|}{Port config.}                       & Single        & Dual        & Single      & Dual         \\ \hline
\multicolumn{1}{|l|}{Rate (ops/s)}                & 500K          & 1M          & 180K        & 190K         \\ \hline
\multicolumn{1}{|l|}{Bottleneck}                  & \multicolumn{2}{c|}{NIC PU} & IB bw       & PCIe bw      \\ \hline
\end{tabular}
}
\caption{NIC throughput of hash lookups and its bottlenecks.}
\label{table:hash_tput}
\end{table}

\smartparagraph{SmartNIC comparison.}  We compare our performance for
hashtable \emph{gets} against StRoM~\cite{strom}, a programmable FPGA-based
SmartNIC. Since we do not have access to a programmable FPGA, we
extract the results from~\cite{strom} for comparison, and report them
in Table~\ref{table:smartnics}. \sys uses the same
experimental settings as before. Our hashtable configuration 
is functionally identical to StRoM's and our client
and server nodes are also connected via back-to-back links.
We can see that
\sys provides lower lookup latencies than StRoM.
StRoM uses a Xilinx Virtex 7 FPGA, which runs at 156.25 MHz, and
incurs at least two PCIe roundtrips to retrieve the key and value. Our
evaluation shows that \sys can provide latency that is in-line with
more expensive SmartNICs.

\begin{figure}[t]
    \centering
    \includegraphics[width=0.5\textwidth]{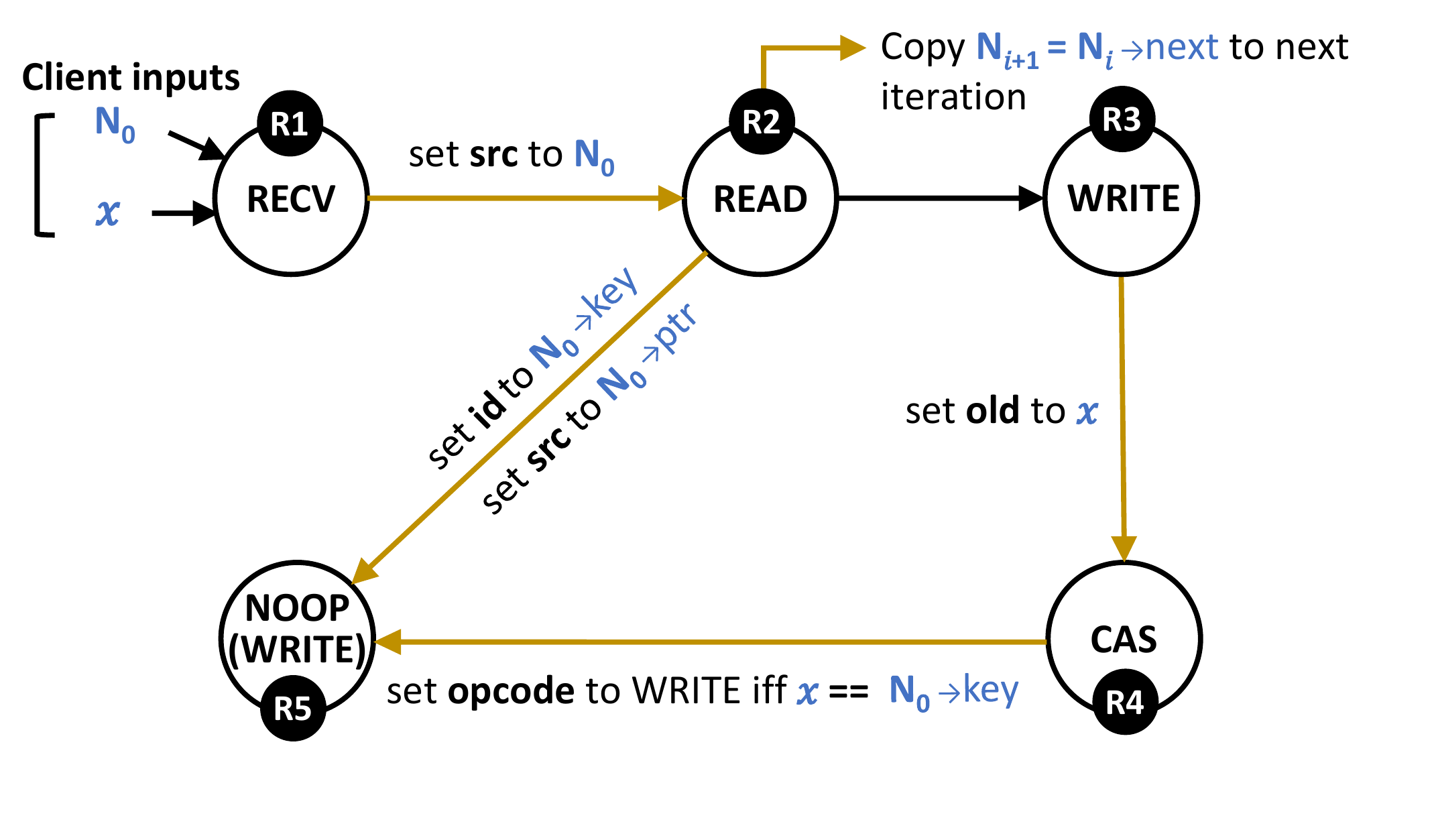}
    \caption{Linked list RDMA program.
    }
    \label{fig:linkedlists_overview}
\end{figure}

\begin{table}[b]
\resizebox{\columnwidth}{!}{%
\begin{tabular}{p{1.5cm}p{1.5cm}p{1.5cm}p{1.5cm}}
\hline
\multicolumn{1}{p{1.5cm}}{\textbf{IO Size}} & \multicolumn{1}{p{1.5cm}}{\textbf{System}} & \multicolumn{1}{p{1.5cm}}{\textbf{Median}} & \multicolumn{1}{p{1.5cm}}{\textbf{$99^{th}$ile}} \\ \hline
\multirow{2}{*}{64 B}                  & \sys                  & 5.7 \us                               & 6.9 \us                                     \\
                                       & StRoM                                & \textasciitilde 7 \us                               & \textasciitilde 7 \us                                     \\ \cline{1-4}
\multirow{2}{*}{4 KB}                  & \sys                  & 6.7 \us                               & 8.4 \us                                     \\
                                       & StRoM                                & \textasciitilde 12 \us                              & \textasciitilde 13 \us                                 
\end{tabular}
}
\caption{Latency comparison of hash \emph{gets}. Results for StRoM obtained from \cite{strom}.}
\label{table:smartnics}
\end{table}

\subsection{Offload: List Traversal}
\label{sec:linkedlist}

Next, we explore another data structure also popularly used in storage systems. We focus on linked lists that store key-value pairs, and evaluate the overhead of traversing them remotely using our offloads. Similar to the previous use-case, we focus on one-sided approaches, as used by FaRM and Pilaf~\cite{farm, pilaf}.

Linked list processing can be decomposed into a \emph{while} loop for
traversing the list and an \emph{if} condition for finding and
returning the key. We describe the implementation of our offload in
Fig. \ref{fig:linkedlists_overview}. The client provides the key $x$
and address of the first node in the list $N_{0}$. A \rread operation
(\circledblack{R2}) is then performed to read the contents of the
first node and update the values for the return operation
(\circledblack{R5}). We also use a \rwrite operation
(\circledblack{R3}) to prepare the CAS operation (\circledblack{R4})
by inserting key $x$ in its \emph{old} field. As an optimization, this
\rwrite can be removed and, instead, $x$ can be inserted directly by
the \rrecv operation. This, however, will need to be done for every
\rcas to be executed and, as such, this approach is limited to smaller
list sizes, since {\rrecv}s can only perform 16 scatters.

For this use-case, we introduce two offload variations. The first, referred to simply as \sys, uses the implementation in Fig. \ref{fig:linkedlists_overview}. The second uses an additional \emph{break} statement between \circledblack{R4} and \circledblack{R5} to exit the loop in order to avoid executing any additional operations.

\begin{figure}[t]
    \centering
    \includegraphics[width=0.5\textwidth]{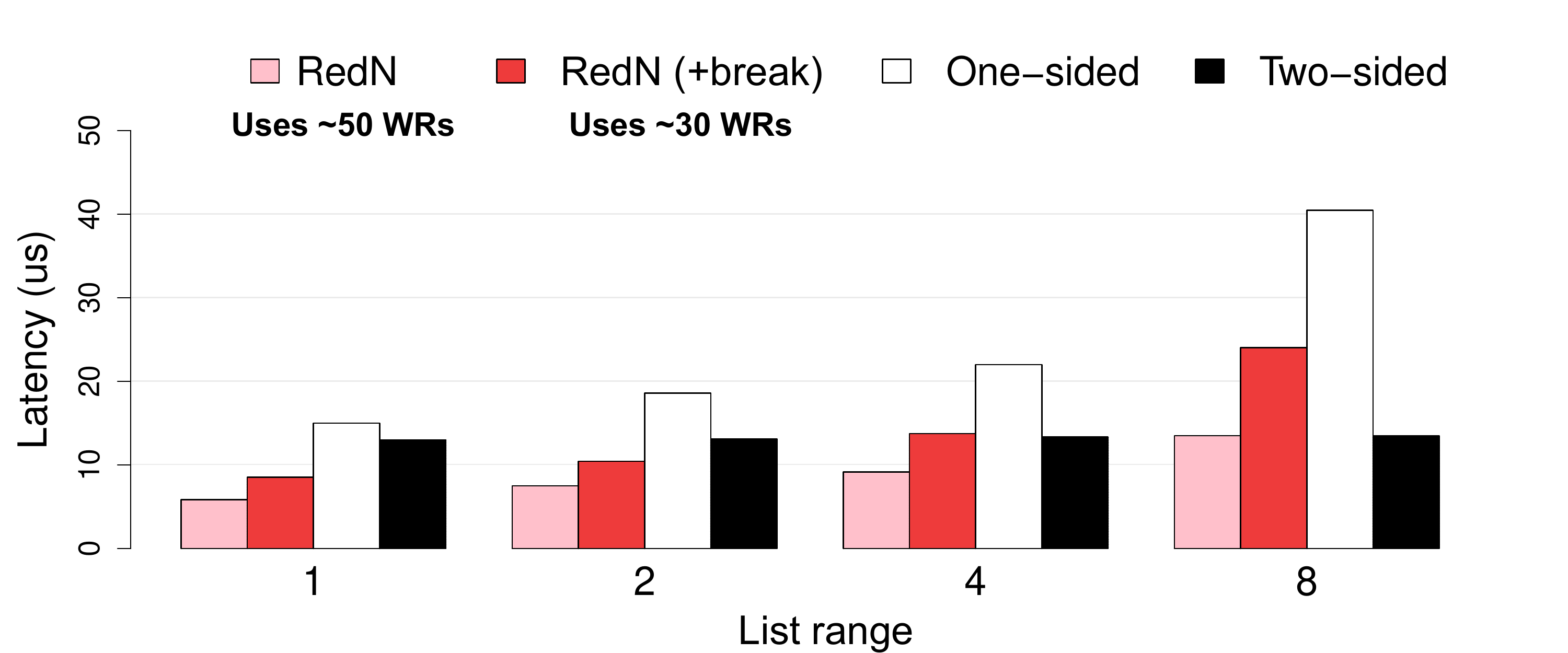}
    \caption{Average latency of walking linked lists.
    }
    \label{fig:linkedlists_lat}
\end{figure}

\subsubsection{Results}
Fig.~\ref{fig:linkedlists_lat} shows the latency of one-sided and
two-sided variants against \sys at various linked list ranges ---
where range represents the highest list element that the key can be randomly placed in.
The size of the list itself is set to a constant value of 8.
We setup the linked list to use key and value sizes of 48 bits and 64 bytes, respectively,
and perform 100k list traversals for each system. The requested key is
chosen at random for each RPC. In the variant labelled ``\sys'', we do
not use \emph{breaks} and assume that all 8 elements of the list need
to be searched.
  \sys outperforms all baselines for all list ranges until 8 
--- providing up to a 2$\times$ improvement. \emph{\sys (+break)} executes a
break statement with each iteration and performs worse than \sys due
to the extra overhead of checking the condition of the
\emph{break}. However, using a break statement increases the offload's
overall efficiency since no unneeded iterations are executed after the
key is found --- using an average of 30 WRs across all experiments. Without breaks,
\sys will need to execute all subsequent iterations even after the key-value
pair is found/returned and it uses more than 65\% more WRs. As such,
while \sys is able to provide better latencies, using a break
statement is more sensible for longer lists.

\subsection{Use Case: Accelerating Memcached}
\label{sec:memcached}

Based on our earlier experience offloading remote data structure traversals,
we set out to see: 1) how effective our aforementioned
techniques are in a real system, and 2) what are the challenges in deploying it in such settings.
Memcached is a key-value store that is often used as a caching service for large-scale storage services. We use a version of Memcached that employs cuckoo hashing~\cite{memc3}. Since Memcached does not natively support RDMA, we modify it with $\sim$700 LoC to integrate RDMA capabilities, allowing the RNIC to register the hash table and storage object memory areas. We also modify the buckets, so that the addresses to the values are stored in big endian --- to match the format used by the WR attributes.
We then use \sys to offload Memcached's \emph{get} requests to allow them to be serviced directly by
the RNIC without CPU involvement. We compare our results to various configurations of Memcached.

To benchmark Memcached, we use the Memtier benchmark, configure it to
use UDP (to reduce TCP overheads for the baselines), and issue 1
million \emph{get} operations using different key-value sizes. To
create a competitive baseline for two-sided approaches, we use
Mellanox's VMA~\cite{libvma}---a kernel-bypass userspace TCP/IP stack
that boosts the performance of sockets-based applications by
intercepting their socket calls and using kernel-bypass to
send/receive data.  We configure VMA in polling-mode to optimize 
for latency. In addition, we also implement a one-sided approach, similar
to the one introduced in section \ref{sec:hash}.

\begin{figure}[t]
    \centering
    \includegraphics[width=0.5\textwidth]{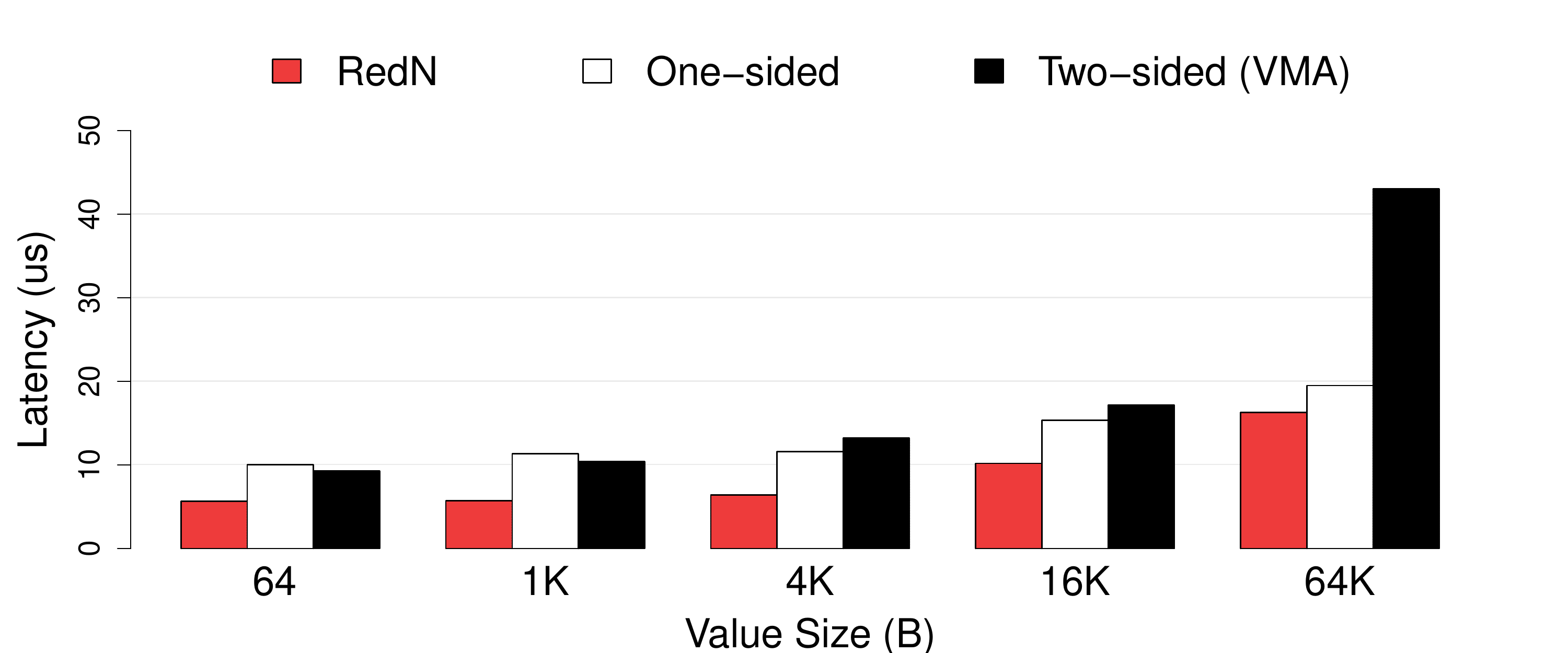}
    \caption{Memcached \emph{get} latencies with different IO sizes.}
    \label{fig:memcached_lat}
\end{figure}

Fig. \ref{fig:memcached_lat}
shows the latency of \emph{gets}. As we can see, \sys's offload for hash
\emph{gets} is up to 1.7$\times$ faster than one-sided and 2.6$\times$ faster than
two-sided. Despite the latter being configured in polling-mode, VMA incurs extra
overhead since it relies on a network stack to process packets. In addition,
to adhere to the sockets API, VMA has to memcpy data from send and
receive buffers, further inflating latencies---which is why it performs
comparatively worse at higher value sizes.

\subsection{Use Case: Performance Isolation}
\label{sec:perf_isolation}

One of the benefits of exposing the latent turing power of RNICs is to
enforce isolation among applications. CPU contention in multi-tenant
and cloud settings can lead to arbitrary context switches, which can,
in turn, inflate average and tail latencies. We explore such a
scenario by sending background traffic to Memcached using one or more
writer (clients). These writers generate \emph{set} RPCs in a closed
loop to load the Memcached service. At the same time, we use a single reader client to generate \emph{get} operations. To  stress CPU resources while minimizing lock contention, each reader/writer is assigned a distinct set of 10K keys, which they use to generate their queries. The keys within each set are accessed by the clients sequentially.

\begin{figure}[t]
    \centering
    \includegraphics[width=0.5\textwidth]{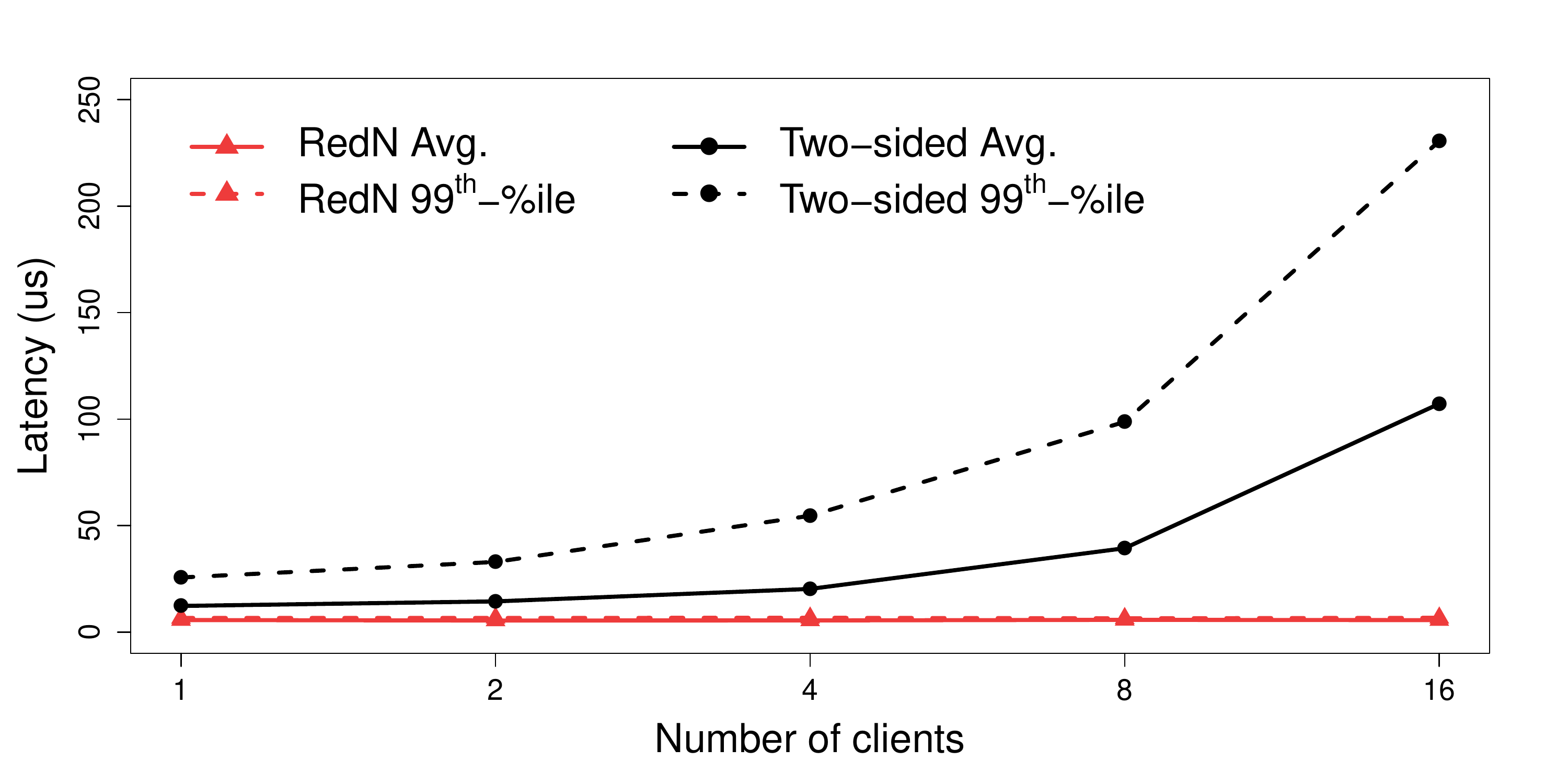}
    \caption{Memcached \emph{get} latencies under hardware contention with varying numbers of writer-clients.}
    \label{fig:memcached_cont}
\end{figure}

We can see in Fig. \ref{fig:memcached_cont} that, as we increase the
\# of writers, both the average and $99^{th}$ percentile latencies for
two-sided increase dramatically. For \sys, CPU contention has no
impact on the performance of the RNIC and
both the average and $99^{th}$ percentiles sit below 7 \us.
At 16 writers, \sys's $99^{th}$ percentile latency is 35$\times$ lower than the baseline.

This indicates that RNIC offloads can also have other useful
effects. Service providers may opt to offload high priority traffic
for more predictable performance or allocate server resources to
tenants to reduce contention.

\subsection{Use Case: Failure Resiliency}
\label{sec:recovery}

We now consider server failures and how failure is affected by
RNICs. Table \ref{table:failure_data} shows failure rates of server
software and hardware components. NICs are much less likely to fail
than software components---NIC annualized failure rate (AFR) is an
order of magnitude lower. Even more importantly, NICs are partially
decoupled from their hosts and can still access memory (or NVM) in the
presence of an OS failure. This means that RNICs are capable of
offloading key system functionality that can allow servers to continue
operating despite OS failures (albeit in a degraded state). To put
this to the test, we conduct a fail-over experiment to explore how
\sys can enhance a service's failure resiliency.

\begin{figure}[t]
    \centering
    \includegraphics[width=0.5\textwidth]{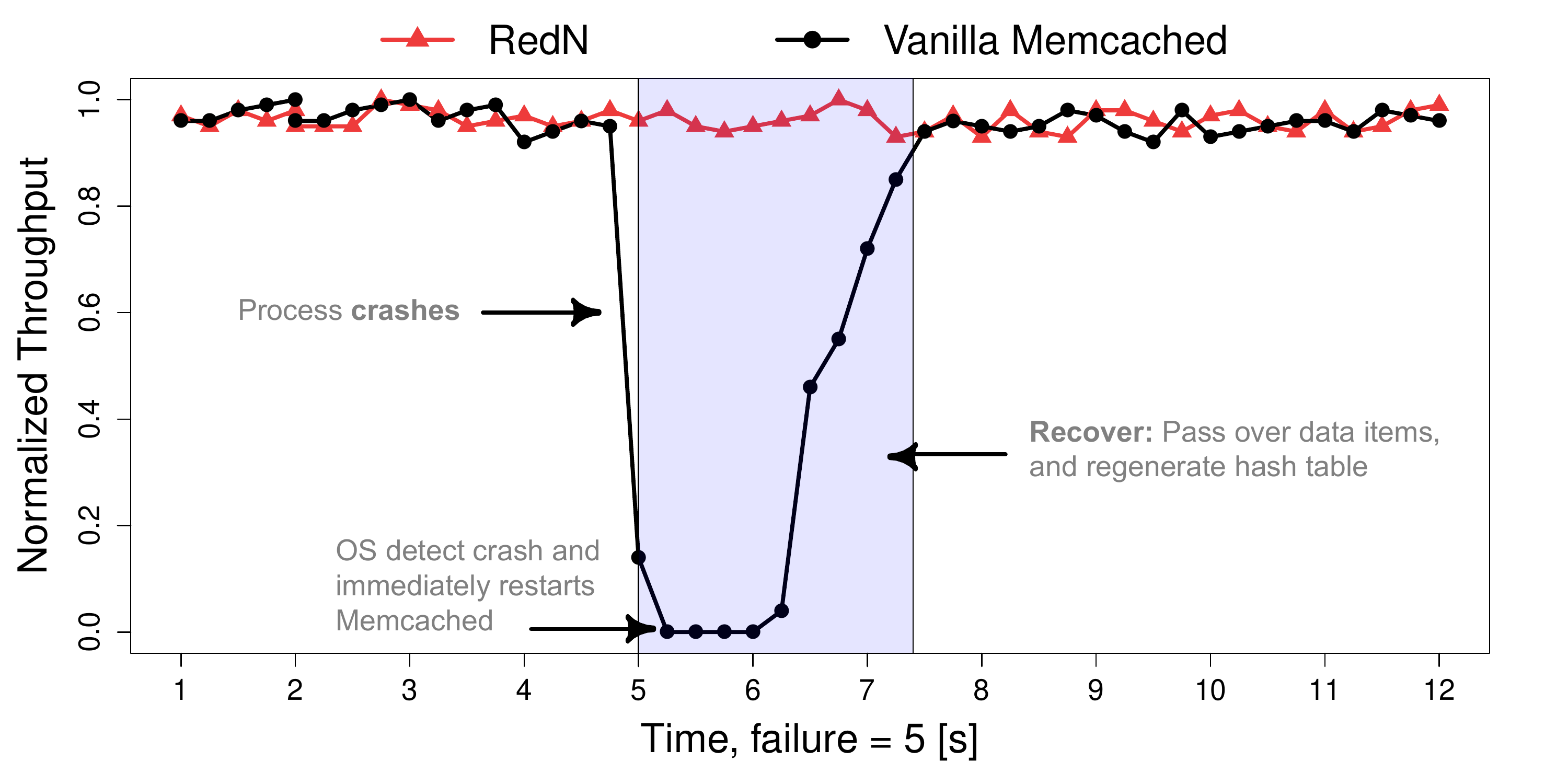}
    \caption{\Sys can survive process crashes and continue serving RPCs via the RNIC without interruption.}
    \label{fig:failover_tput}
\end{figure}

\smartparagraph{Process crashes.} We look into how we can allow an
RNIC to continue serving RPCs after a Memcached instance crashes. We
find that this is not simple in practice. RNICs access many resources
in application memory (\eg queues, doorbell records, \etc) that are
required for functionality. If the process hosting these resources
crashes, the memory belonging to these components will be
automatically freed by the operating system resulting in termination
of the RDMA program. To counteract this, we use~\cite{rdmaforks} forks
to create an empty hull parent for hosting RDMA resources and then
allow Memcached to run as a child process. Linux systems do not free
the resources of a crashed child until the parent also terminates. As
such, keeping the RDMA resources tied to an empty process allows us to
continue operating in spite of application failures. We run an
experiment (timeline shown in Fig.~\ref{fig:failover_tput}) where we
send \emph{get} queries to a single instance of Memcached and then
simply kill Memcached during the run. The OS
detects the application's termination and immediately restarts
it. Despite this, we can see that a vanilla Memcached instance will
take at least 1 second to bootstrap, and 1.25 additional seconds to
build its metadata and hashtables. With \sys, no service disruption is
experienced and \emph{get} queries continue to be issued without
recovery time.

\smartparagraph{OS failure.} We also programmatically induce a kernel
panic using \textsf{sysctl}, freezing the system.  This is a simpler
case than process crashes, since we no longer have to worry about the
OS freeing RDMA resources. For brevity, we do not show these results,
but we experimentally verified that \sys offloads continue operating
in the presence of an OS crash.

\begin{table}
\centering
\begin{tabular}{|c|c|c|c|}  \hline
\multicolumn{1}{|l|}{\textbf{Component}} & \textbf{AFR} & \textbf{MTTF} & \textbf{Reliability} \\  \cline{1-4}
OS                                     & 41.9\%       & 20,906        & 99\%              \\ \cline{1-4}
DRAM                                   & 39.5\%       & 22,177        & 99\%              \\ \cline{1-4}
NIC                                    & 1.00\%       & 876,000       & 99.99\%              \\ \cline{1-4}
NVM                                    & $<$ 1.00\% & 2 million     & 99.99\% \\ \cline{1-4}
\end{tabular}
\caption{Failure rates of different server components~\cite{poke2015dare, optane}. AFR means annualized failure rate, whereas MTTF stands for mean time to failure and is expressed in hours. RNICs can still access memory even in the presence of an OS failure.}
\label{table:failure_data}
\end{table}

\section{Discussion}
\label{sec:discussion}

\smartparagraph{Client scalability.} \Sys requires servers to manage
at least two WQs per client, which is not higher than other RDMA
systems.
\Sys can still introduce scalability challenges with thousands of
clients since RNIC cache is limited. However, Mellanox's
dynamically-connected (DC) transport service \cite{dctransport}, which
allows unused connections to be recycled, can circumvent many such
scalability limits.

\smartparagraph{Offload for sockets-based applications.}
Protocols such as rsocket~\cite{rsocket} can be used to transparently
convert sockets-based applications to use RDMA, making them
possible targets for \sys. Although rsocket does not support
popular system calls, such as \textsf{epoll}, other extensions have
been proposed~\cite{socksdirect} that support a more comprehensive
list of system calls and were shown to work with applications like
Memcached and Redis.

\smartparagraph{Intel RNICs.}
Next-generation Intel RNICs are expected to support atomic verbs, such as CAS---which \sys uses to implement conditionals. To control when WRs can
be fetched by the NIC, Intel uses a validity bit in each WR
header. This bit can be dynamically modified via an RDMA operation to
mimic {\renable}. However, there is no equivalent for the {\rwait}
primitive, meaning that clients cannot trigger a pre-posted chain. One
possible workaround for this is to use another PCIe device on the
server to issue a doorbell to the RNIC, allowing the WR chain to be
triggered. We leave the exploration of such techniques as future work.

\smartparagraph{Insights for next-generation RNICs.}  Our
experience with \sys has shown that keeping WRs in server
memory (to allow them to be modified by other RDMA verbs) is a key
bottleneck. If the NIC's cache was made directly accessible via RDMA,
WRs can be pre-fetched in advance and unnecessary PCIe round-trips on the critical path can be avoided.
We hope future RNICs will support such features. %

\section{Conclusion}

We show that, in spite of appearances, commodity RDMA NICs are
Turing-complete and capable of performing complex offloads without
\emph{any} hardware modifications. We take this insight and explore
the feasibility and performance of these offloads. We find that, using
a commodity RNIC, we can achieve up to 2.6$\times$ and 35$\times$
speed-up versus state-of-the-art RDMA approaches, for key-value get
operations under uncontended and contended settings, respectively, while
allowing applications to gain failure resiliency to OS and process
crashes. We believe that this work opens the door for a wide variety
of innovations in RNIC offloading which, in turn, can help guide the evolution of the RDMA standard.\\
\sys is available at \url{https://redn.io}.

\paragraph{Acknowledgements.} This work has received funding from the
European Research Council (ERC) under the European Union’s Horizon
2020 research and innovation programme (grant agreement No. 770889),
as well as NSF grant 1751231. 
Waleed Reda was supported by a
fellowship from the Erasmus Mundus Joint Doctorate in Distributed
Computing (EMJD-DC), funded by the European Commission (EACEA) (FPA
2012-0030).
We would like to thank Gerald Q. Maguire Jr. and our anonymous
reviewers for their comments and feedback as well as Jasmine Murcia.
Thanks also go to our shepherd Ang Chen.
\\

\label{lastpage}

{\bibliographystyle{abbrv}
\bibliography{references}}
\begin{appendices}

\onecolumn
\begin{table*}[bp]
\centering
\begin{tabular}{|m{3cm}|m{3.5cm}|m{7cm}|}
\cline{1-3}
\textbf{Addressing mode}    & \textbf{x86 syntax}                             & \textbf{\sys equivalent} \\ \cline{1-3}

Immediate & \texttt{mov} $R_{dst}$, C                    & 
\raisebox{-.5\height}{\includegraphics[height=8mm, trim=0 0 0 -10]{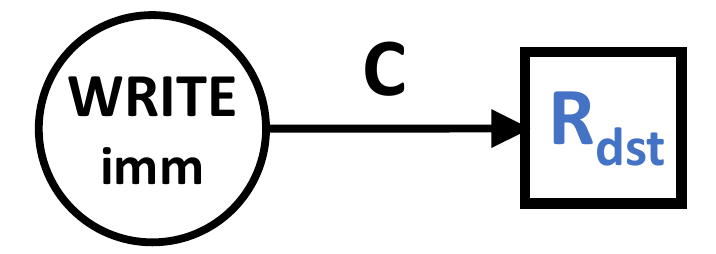}}

\\ \cline{1-3}
Indirect   & \texttt{mov} $R_{dst}$, {[}$R_{src}${]}  &

\raisebox{-.5\height}{\includegraphics[height=14mm, trim=0 0 0 -10]{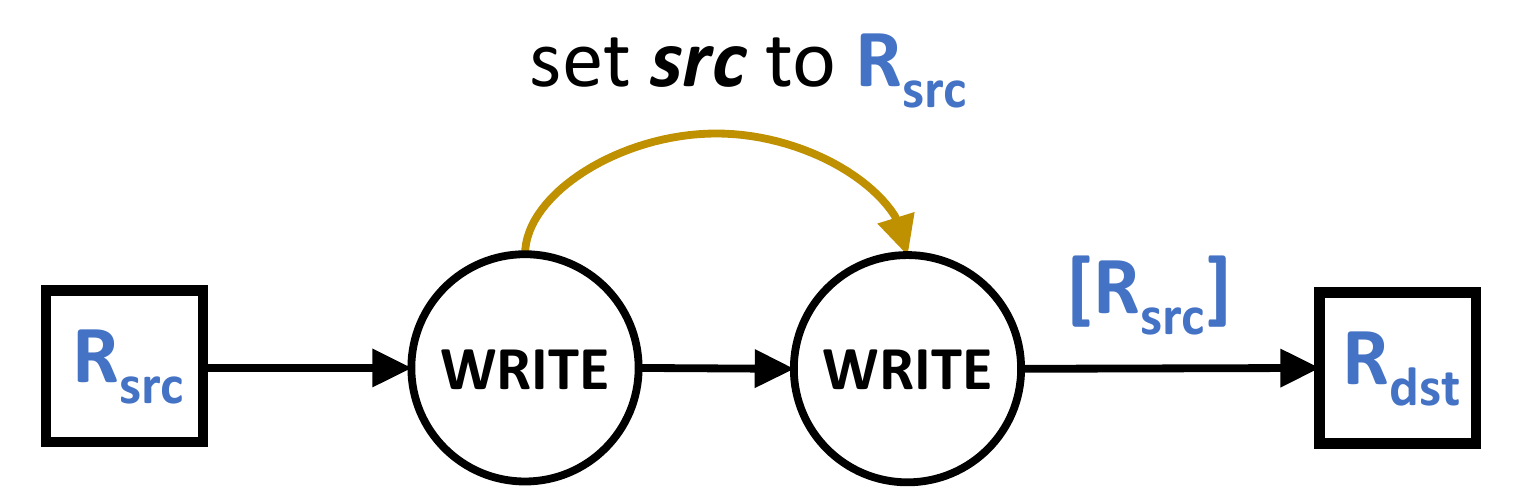}}

\\ \cline{1-3}
Indexed  & \texttt{mov} $R_{dst}$, {[}$R_{src}$ + $R_{off}${]} & 
\raisebox{-.5\height}{\includegraphics[height=20mm, trim=0 0 0 -10]{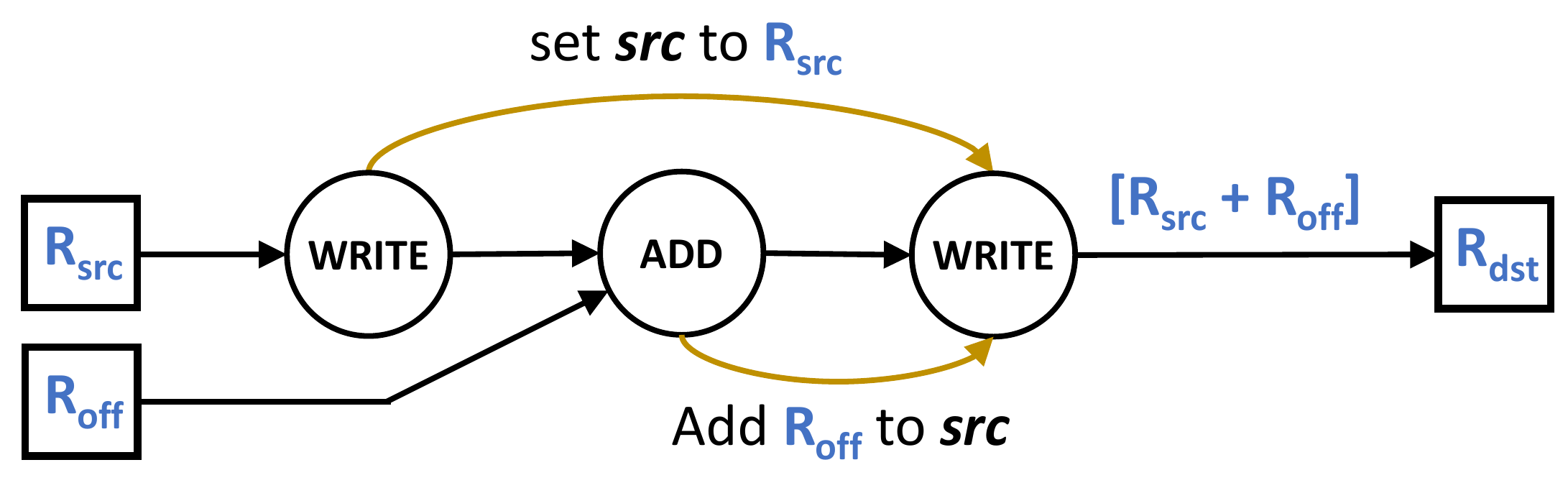}}

\\ \cline{1-3}
\end{tabular}
\captionof{table}{Addressing modes for the x86 mov instruction and their RDMA implementation in \sys.}
\label{table:mov}
\end{table*}

\begin{multicols}{2}

\section{Turing completeness sketch}
\label{appendix:turing}

To show that RDMA is turing complete, we need to establish that RDMA has the following three properties:

\begin{enumerate}[noitemsep]
\item Can read/write arbitrary amounts of
  memory.

\item Has conditional branching (e.g., if \& else
  statements).

\item Allows nontermination.
  
\end{enumerate}

Our paper already demonstrates that these properties can be satisfied using our constructs but, for completeness, we also analogize our system with \texttt{x86} assembly instructions that have been proven to be capable of simulating a Turing machine. Dolan~\cite{dolan2013mov} demonstrated that this is in fact possible using just the \texttt{x86} \texttt{mov} instruction. As such, we need to prove that RDMA has sufficient expressive power to emulate the \texttt{mov} instruction.

\subsection{Emulating the \texttt{x86 mov} instruction}
To provide an RDMA implementation for \texttt{mov}, we first need to consider the different addressing modes used by Dolan~\cite{dolan2013mov} to simulate a Turing machine. The addressing mode describes how a memory location is specified in the \texttt{mov} operands.

Table \ref{table:mov} shows a list of all required addressing modes, their \texttt{x86} syntax, and one possible implementation for each with RDMA. $R$ operands denote registers but, since RDMA operations can only perform memory-to-memory transfers, we assume these registers are stored in memory. For simplicity, we only focus on \texttt{mov} instructions used to perform \emph{loads} but note that \emph{stores} can be implemented in a similar manner.

For \emph{immediate} addressing, the operand is part of the instruction and is passed directly to register $R_{dst}$. This can be implemented simply using an {\rwriteimm} which takes a constant in its \emph{immediate} parameter and writes it to a specified memory location (register $R_{dst}$ in this case). To perform more complex operations, \emph{indirect} allows \texttt{mov} to use the value of the operand as a memory address. This enables the dynamic modification of the address at runtime, since it depends on the contents of the register when the instruction is executed. To implement this, we use two write operations with doorbell ordering (refer to \S\ref{sec:consistency} for a discussion of our ordering modes). The first {\rwrite} changes the \emph{source address} attribute of the second {\rwrite} operation to the value in register $R_{src}$. This allows the second {\rwrite} operation to write to register $R_{dst}$ using the value at the memory address pointed to by $R_{src}$. Lastly, \emph{indexed} addressing allows us to add an offset ($R_{off}$) to the address in register $R_{src}$. This can be done by simply performing an RDMA {\radd} operation between the two writes with doorbell ordering, in order to add the offset register value $R_{off}$ to $R_{src}$. This allows us to finally write the value $[R_{src} + R_{off}]$ to $R_{dst}$. With these three implementations, we showcase that RDMA can in fact emulate all the required \texttt{mov} instruction variants.

\subsection{Allowing nontermination}
To simulate a real Turing machine, we need to also satisfy the code nontermination requirement. In the \texttt{x86} architecture, this can be achieved via an unconditional jump~\cite{dolan2013mov} that loops back to the start of the program. For RDMA, this can also be achieved by having the CPU re-post the WRs after they are executed. While this is sufficient for Turing completeness it, nevertheless, wastes additional CPU cycles and can also impact latency if CPU cores are busy or unable to keep up with WR execution. As an alternative, \Sys provides a way to loop back without any CPU interaction by relying on {\rwait} and {\renable} to recycle RDMA WRs (as described in \S\ref{sec:recycling}). Regardless of which approach is employed, RDMA is capable of performing an unconditional jump to the beginning of the program. This means that we can emulate all \texttt{x86} instructions used by Dolan~\cite{dolan2013mov} for simulating a Turing machine.

\end{multicols}

\end{appendices}

\end{document}